\documentclass[lettersize,journal]{IEEEtran}
\usepackage{amsmath,amsfonts}
\usepackage{algorithmic}
\usepackage{algorithm}
\usepackage{array}
\usepackage[caption=false,font=normalsize,labelfont=sf,textfont=sf]{subfig}
\usepackage{textcomp}
\usepackage{stfloats}
\usepackage{url}
\usepackage{verbatim}
\usepackage{graphicx}
\usepackage{cite}
\usepackage{lineno}
\usepackage[colorlinks,allcolors=blue]{hyperref}
\usepackage{amstext}
\usepackage{amssymb}
\usepackage{color}
\usepackage{booktabs}
\usepackage{multirow}
\usepackage{mathtools}
\usepackage[all]{xy}
\usepackage[T1]{fontenc}

\newtheorem{defi}{Definition}

\graphicspath{{images/}}
\DeclareGraphicsExtensions{.pdf,.png,.jpg}

\begin{document}

\title{Test Adequacy for Metamorphic Testing: Criteria, Measurement, and Implication}

\author{An Fu, Chang-ai Sun,~\IEEEmembership{Senior Member,~IEEE,} Jiaming Zhang, and Huai Liu,~\IEEEmembership{Senior Member,~IEEE}

\thanks{Manuscript received 4 February 2024; revised XX XXXX 2024. This work was supported by the National Natural Science Foundation of China (Grant Nos. 62272037 and 61872039), the Beijing Municipal Natural Science Foundation (Grant No. 4162040), the Aeronautical Science Foundation of China (Grant No. 2016ZD74004), the Fundamental Research Funds for the Central Universities (Grant No. FRF-GF-17-B29), and ARC DP210102447. \textit{(Corresponding author: Chang-ai Sun.)}}
\thanks{An Fu, Chang-ai Sun and Jiaming Zhang are with the Department of Computer Science and Technology, University of Science and Technology Beijing, Beijing 100083, China (e-mail: anfu@xs.ustb.edu.cn, casun@ustb.edu.cn, radon@xs.ustb.edu.cn). }
\thanks{Huai Liu is with the Department of Computing Technologies, Swinburne University of Technology, Melbourne, Australia (e-mail: hliu@swin.edu.au). }
}

\markboth{IEEE XXXXX,~Vol.~00, No.~0, February~2024}%
{Fu \MakeLowercase{\textit{et al.}}: Test Adequacy for Metamorphic Testing: Criteria, Measurement, and Implication}

\IEEEpubid{0000--0000/00\$00.00~\copyright~2024 IEEE}

\maketitle

\begin{abstract}
Metamorphic testing (MT) is a simple yet effective technique to alleviate the oracle problem in software testing. The underlying idea of MT is to test a software system by checking whether metamorphic relations (MRs) hold among multiple test inputs (including source and follow-up inputs) and the actual output of their executions. Since MRs and source inputs are two essential components of MT, considerable efforts have been made to examine the systematic identification of MRs and the effective generation of source inputs, which has greatly enriched the fundamental theory of MT since its invention. However, few studies have investigated the test adequacy assessment issue of MT, which hinders the objective measurement of MT's test quality as well as the effective construction of test suites.
Although in the context of traditional software testing, there exist a number of test adequacy criteria that specify testing requirements to constitute an adequate test from various perspectives, they are not in line with MT's focus which is to test the software under testing (SUT) from the perspective of necessary properties.
In this paper, we proposed a new set of criteria that specifies testing requirements from the perspective of necessary properties satisfied by the SUT, and designed a test adequacy measurement that evaluates the degree of adequacy based on both MRs and source inputs.
A series of experiments involving seven programs and five coverage criteria were conducted to evaluate the performance of the proposed criteria.
The experimental results have shown that the proposed measurement can effectively indicate the fault detection effectiveness of test suites, i.e., test suites with increased test adequacy usually exhibit higher effectiveness in fault detection.
Our work made an attempt to assess the test adequacy of MT from a new perspective, and our criteria and measurement provide a new approach to evaluate the test quality of MT and provide guidelines for constructing effective test suites of MT.
\end{abstract}


\section{Introduction}\label{sec:intr}

\IEEEPARstart{S}{oftware} systems have been playing a vital role in human daily life, and their quality assurance is an important issue.
However, at the same time, software failures have caused severe problems; for example, it was reported that in 2017 alone, software failures resulted in \$1.7 trillion in losses and impacted half of the world's population (3.7 billion people)~\cite{tricentis2017}.
As a prominent technique to ensure the quality of software systems, software testing aims at revealing faults of the software under test (SUT) by executing test cases and evaluating the correctness of software outputs.
The majority of existing techniques, explicitly or implicitly, assume the availability of test oracles for checking the correctness of the outputs, that is, test oracles can be precisely constructed without much effort. However, in many situations, a test oracle does not exist, or is too expensive to be acquired and applied, which is referred to as the \emph{oracle problem}~\cite{barr2015oracle, patel2018mapping}. The oracle problem significantly restricts the applicability of testing techniques that rely on test oracles.

\IEEEpubidadjcol

Metamorphic testing (MT)~\cite{chen1998metamorphic} is a promising technique that effectively alleviates the oracle problem. Rather than directly checking the correctness of software outputs, MT checks whether some necessary properties hold among multiple inputs and their corresponding outputs. These necessary properties are known as \emph{metamorphic relations (MRs)} which are commonly identified from the software specification. Metamorphic testing is performed by executing multiple inputs and then checking whether the corresponding multiple outputs satisfy some expected relation expressed by an MR. Once an MR is not satisfied, SUT is said to be faulty.
Apparently, MT provides a new perspective on revealing software faults by involving multiple executions of SUT and verifying their corresponding outputs using MRs, which alleviates the need for expected outputs. So far, MT has been widely adopted in various application domains and was demonstrated effective at alleviating the oracle problem, including autonomous driving systems~\cite{tian2018deeptest}, compilers~\cite{xiao2022metamorphic}, encryption programs~\cite{sun2014property}, citation database systems~\cite{zhou2019TSE}, question answering models~\cite{yuan2021perception}, object detection systems~\cite{zhang2021deep}, and concurrent programs~\cite{sun2024ConMT}. Being widely accepted by the industry, MT was added into an ISO/IEC/IEEE standard~\cite{IEEE2021Standard}.

With the increasing popularity of MT, substantial efforts have been made in various aspects as well to contribute to the fundamental theory of MT. Since MRs and test inputs (more precisely, source test inputs) are two key components of MT, the identification of MRs and the generation of source test inputs have been two hot research topics of MT. So far, existing studies have extensively investigated different mechanisms to identify SUT's MRs~\cite{rahamn2021using, qiu2022theoretical, sun2016mumt}, and it has been reported that a large number of MRs can be systematically identified~\cite{chen2016metric, sun2021metricplus, segura2022automated}. In addition, some studies have also examined the characteristic of effective MRs and proposed some qualitative guidelines for selecting effective MRs~\cite{liu2014effectively, mayer2006empirical}. Apart from the identification of MRs, a variety of source test input generation approaches have been proposed~\cite{sun2022path, alatawi2016source}, and studies have investigated their impact on the fault detection effectiveness of MT~\cite{barus2016impact, saha2018fault}.

In contrast to the considerable effort devoted to the identification of MRs and the generation of source test inputs, little attention was given to the test adequacy assessment of MT.
Test adequacy is a key factor for testing in general, which not only indicates the quality of testing performed on SUT but also provides a rule that can help decide reasonably when and where to stop testing~\cite{zhu1997software}.
The basic principle of assessing test adequacy is to evaluate the extent to which test cases employed in testing satisfy some test requirements specified by a \emph{test adequacy criterion}; these test requirements are expected to be fully achieved to constitute a ``thorough'' test.
In this regard, test requirements specified by a test adequacy criterion provide intuitive clues for the selection of test cases to achieve high quality of testing. Despite these fundamental roles, most studies on MT were conducted based on a set of arbitrarily constructed MRs and test inputs without considering their test adequacy.
Although the result of these studies usually demonstrated that MT can effectively detect software defects,
such results failed to give an insight how thoroughly MT has tested the SUT, which is critical for giving an in-depth evaluation of the overall quality.
In practice, a testing activity pursues the detection of as many software faults as possible within limited resources. In this regard, test adequacy shall be a fundamental and essential aspect for any testing technique, including MT. Unfortunately, such adequacy criteria and measurement are absent for MT, as manifested by the ISO/IEC/IEEE international standard for software testing\cite{IEEE2021Standard}.

Some preliminary studies attempted to tackle this important issue~\cite{ding2017application, lascu2021dreaming}.
Their underlying idea was to leverage existing test adequacy criteria for conventional testing techniques\footnote{Here, we call those testing techniques that rely on test oracles as conventional testing techniques.} to assess the test adequacy of MT.
Unfortunately, \emph{they failed to take into consideration the inherent characteristic of MT, leading to a controversial estimation of MT's test quality.}
Specifically speaking, MT checks whether the SUT satisfies some necessary properties; as the carrier of these necessary properties, MRs play a vital role in MT and strongly affect its fault detection capability.
Meanwhile, existing test adequacy criteria are unaware of MRs. For instance, the program structure-based criteria~\cite{mathur2013foun} specify testing requirements in terms of elementary components of the program, specification-based criteria~\cite{bernot1991software} specify testing requirements in terms of the features of software specification, fault-based criteria~\cite{demillo1989test} specify testing requirements in terms of the software faults, and the operational profile-based criteria~\cite{miranda2017ass} specify testing requirements in terms of the usage profile of a software system. Their unawareness of software necessary properties calls for new adequacy criteria for MT.

In this paper, we propose a new set of test adequacy criteria and a measurement for MT. Since both MRs and source inputs have indispensable contributions to the quality of MT, they shall be fully considered when assessing test adequacy.
Accordingly, the underlying intuition behind our criteria is to specify test requirements that make use of both MRs and source inputs collectively to constitute a thorough testing.
On the one hand, test requirements for source inputs should enable MT to exercise various parts of the SUT as much as possible; on the other hand, test requirements for MRs should enable MT to verify the test results of SUT using various necessary properties as much as possible.
Considering that there exist many well-developed test adequacy criteria based on test inputs,
our criteria reuses an existing test case adequacy criterion to specify test requirements for source inputs. Since MRs not only associate with source inputs to generate follow-up inputs but also decide whether SUT is faulty according to the relation of outputs (hereafter referred to as output sub-relation in the rest of this paper), which are two factors that impact the test quality of MT, our criteria specify requirements regarding the way of association between MRs and source inputs.
With these ideas in mind, we proposed the so-called $k$-MR coverage criterion and a corresponding test adequacy measurement.

The main contributions of the paper are:
\begin{itemize}
  \item Test adequacy criteria for MT are proposed, which specify test requirements from both perspectives of high-level SUT's necessary properties and low-level concrete test cases.
  \item A test adequacy measurement is designed, which quantifies the degree of test adequacy based on both the MRs and source inputs employed in MT. The proposed test adequacy criteria and test adequacy measurement fill in the gap in coverage calculation for MT.
  \item A series of empirical studies, involving seven programs and five test case coverage criteria, were conducted to comprehensively evaluate the effectiveness of the proposed criteria and the measurement. The empirical result endorses the usefulness of criteria and the measurement for improving the test effectiveness of MT.
\end{itemize}

The rest of this paper is organized as follows: Section~\ref{sec:back} introduces the basic concepts of MT and the important notions of test adequacy assessment. Section~\ref{sec:crit} describes the motivation and the test adequacy criteria. Section~\ref{sec:sett} describes the settings of our empirical study. Sections~\ref{sec:empi} and~\ref{sec:realFault} reports and discusses the results of empirical study. Section~\ref{sec:rela} discusses related work. Finally, Section~\ref{sec:conclusion} concludes this paper.

\section{Background}
\label{sec:back}

This section revisits MT and test adequacy criteria.

\subsection{Metamorphic Testing}
\label{sec:back-mt}

MT was proposed by Tsong Yueh Chen in late 1990's to provide a simple yet effective mechanism to detect the defects hidden in an SUT when the test oracle is unavailable.
The underlying intuition of MT is as follows:
the functions implemented by the SUT have some essential properties that reflect some inner relations among multiple inputs and their expected outputs,
and such relations should hold in the SUT; otherwise, the SUT must be faulty.
By executing the SUT using multiple test inputs and checking whether these relations, commonly known as metamorphic relations (MRs), are satisfied among the test inputs and their execution outputs, the SUT is tested without the need for oracles of individual test inputs. During this process, MT requires collaboration between MRs and test inputs for testing. We next present some essential notions of MT.

Given a program \emph{P} that is implemented according to a function (or an algorithm) \emph{f}:\emph{X}$\rightarrow$\emph{Y}, where \emph{X} denotes the domain of \emph{f} and \emph{Y} denotes the range of \emph{f}.
An MR of \emph{P} refers to a necessary property of \emph{f} in relation to at least two inputs of \emph{f} and their corresponding images under \emph{f}, expressed as $\mathcal{R}$$\subseteq$\emph{X}$^{n}\times$\emph{Y}$^{n}$, where \emph{X}$^{n}$ denotes the $n$-ary Cartesian power of set \emph{X} (and analogous to \emph{Y}$^{n}$).
For simplicity, we use $\mathcal{R}(t_{1}, t_{2}, \dots, t_{n}, f(t_{1}), f(t_{2}), \dots, f(t_{n}))$ to indicate that $\langle t_{1}, t_{2}, \dots, t_{n}, f(t_{1}), f(t_{2}), \dots, f(t_{n}) \rangle \in \mathcal{R}$~\cite{chen2018metamorphic}, where $t_{1}, t_{2}, \dots, t_{n} \in$\emph{X} and $f(t_{1}), f(t_{2}), \dots, f(t_{n})\in$\emph{Y}.
For all 2\emph{n}-tuples in $\mathcal{R}$, there exists a $k$~($1\leq k<n$) such that within each tuple $t_{k+1}$, $t_{k+2}$, $\dots$, $t_{n}$ can be constructed from $t_{1}$, $t_{2}$, $\dots$, $t_{k}$ and their corresponding images $f(t_{1})$, $f(t_{2})$, $\dots$, $f(t_{k})$ based on $\mathcal{R}$. Specifically:
\begin{itemize}
  \item Each $t_{i}$ $(1\leqslant i\leqslant k)$ is referred to as a \emph{source input}.
  \item Each $t_{i}$ $(k+1\leqslant i\leqslant n)$ is referred to as a \emph{follow-up input}.
  \item Sequence $\langle t_{1},\dots,t_{k},t_{k+1},\dots,t_{n}\rangle$ is referred to as a \emph{metamorphic group of test inputs}~\cite{chen2018metamorphic} for $\mathcal{R}$, or simply a metamorphic group (MG) for $\mathcal{R}$.
\end{itemize}

When the construction of follow-up inputs is unrelated to the execution outputs of source inputs, an MR can be considered as having two parts: an \emph{input subrelation} $R_{\rm{in}}$ and an \emph{output subrelation} $R_{\rm{out}}$, denoted as $\mathcal{R}=\langle R_{\rm{in}},R_{\rm{out}}\rangle$.


The general steps for implementing MT are as follows:
\begin{enumerate}
  \item For a given function or algorithm, some necessary properties among multiple inputs and their outputs are identified and represented as MRs.
  \item Source inputs are generated using a conventional test case generation technique, and are executed to obtain their corresponding outputs.
  \item Follow-up inputs are generated from the source inputs (and their outputs if necessary) in Step (2) based on the MRs defined in Step (1), and are executed to obtain their corresponding outputs.
  \item The outputs of source and follow-up inputs are examined with reference to the MRs in Step (1). If some MRs are violated, then the SUT is faulty; otherwise, the SUT passes the test.
\end{enumerate}

Commonly, MGs are considered as the unit of test execution and test result verification. For an MG, after executing its source and follow-up test cases, the execution outputs are verified according to the output subrelation of an MR.

\subsection{Test Adequacy Criteria}

Exhaustive testing ascertains whether the SUT contains defects~\cite{poole1975debugging}. However, this approach is infeasible in practice. Accordingly, software testers turn to sample a finite number of inputs from all possible ones to form a set of test cases, aiming to test SUT as thoroughly as possible. A significant theory of test case selection was proposed by Goodenough and Gerhart~\cite{goodenough1975toward} in 1975. Their theory defines a ``thorough'' test set to be one fully satisfying a carefully designed test criterion that describes what program properties must be exercised in testing to constitute a thorough test. Their theory on the adequacy criteria provided a substantial reference value, which motivated researchers to study the desired properties of criteria~\cite{weyuker1986axio, zhu1993test} and to investigate practically applicable criteria from various aspects~\cite{demillo1978hints, frankl1988app}. The past few decades have seen the rapid development of test adequacy criteria, and a large number of criteria have been proposed from different aspects, e.g., program structure-based criteria~\cite{mathur2013foun}, specification-based criteria~\cite{bernot1991software}, fault-based criteria~\cite{demillo1989test}, and operational profile-based criteria~\cite{miranda2017ass}.
At the same time, test adequacy criteria have received growing attention from the software industry and gradually become an indispensable step of software quality assurance~\cite{shi2019industry, bogner2021industry}. In general, a test adequacy criterion plays important roles in the following aspects:
\begin{itemize}
  \item A test adequacy criterion can serve as a measurement to assess test quality. In this regard, a criterion is regarded as a function that accepts a program $p$ to be tested, a specification $s$, and a test set $t$ and outputs a real number $r$ in the range of $[0,1]$ which represents the degree of adequacy~\cite{zhu1997software}, expressed as $C:P\times S \times T \rightarrow [0,1]$, where $P$ denotes a set of programs, $S$ denotes a set of specifications, and $T$ denotes the power set of the set of all possible inputs of programs in $P$.
      The greater the value of $r$ the higher the adequacy.
      In particular, when the value of adequacy degree is restricted to the set $\{0, 1\}$, where ``0'' denotes \emph{false} and ``1'' denotes \emph{true}, a test adequacy criterion can serve as an indicator that tells a software tester whether or not enough testing has been carried out and it is reasonable to terminate testing.
  \item A test adequacy criterion can serve as a set of guidelines to select test cases. On the one hand, a criterion specifies the requirements on test cases that are capable of adequately testing a target program.
      Accordingly, given a program under test and its specification, one can follow these requirements to produce various sets of test cases that fully satisfy these requirements, each of which is considered to be an adequate test set.
      In this regard, a criterion is regarded as a function that accepts a program $p$ and a specification $s$, and outputs a set that consists of adequate test sets~\cite{zhu1997software}, expressed as $P\times S\rightarrow 2^T$.
      On the other hand, a criterion tells a tester what needs to be observed during the testing process such that it is able to determine whether testing is adequate according to a criterion.
\end{itemize}

In this paper, we focus on the role of test quality measurement played by test adequacy criteria in the context of MT.

\section{Adequacy Criteria and Measurement for MT}
\label{sec:crit}


\subsection{Motivation}

\begin{figure}[t]
    \centering
    \includegraphics[width=0.9\columnwidth]{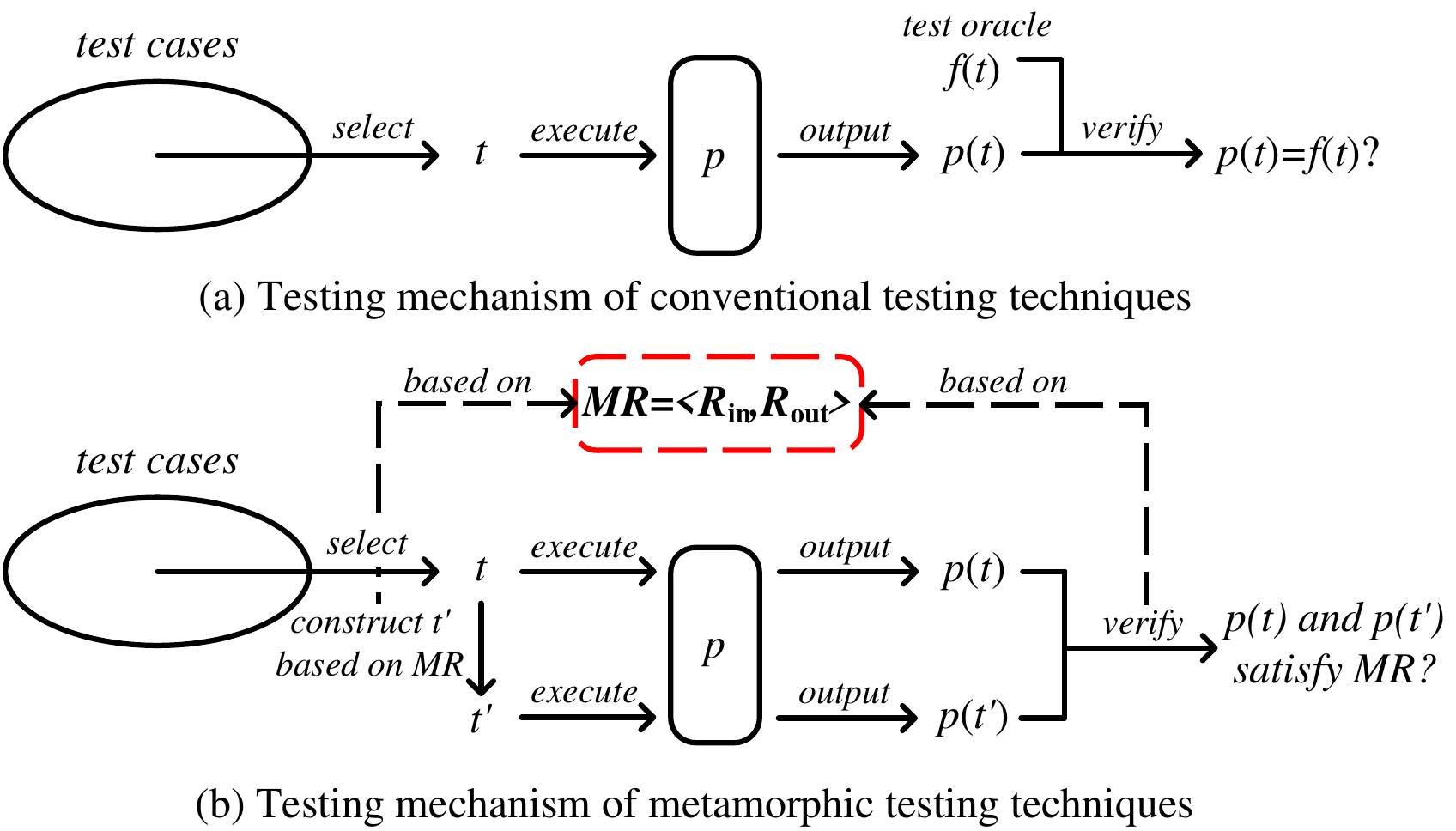}
    \caption{Comparison between MT and conventional testing techniques. $p(t)$ denotes the actual output of executing input $t$ on $p$, $f(t)$ denotes the test oracle of $p(t)$. Figure 1(a) illustrates the test result verification mechanism of conventional testing techniques; figure 1(b) illustrates the test result verification mechanism of MT using an MR in relation two test inputs.}
    \label{fig:verf-mech}
\end{figure}

Fig.~\ref{fig:verf-mech} shows the fundamental difference in testing mechanism between MT and conventional testing techniques.
Given a program $p$ that implements a function $f$ whose domain and range are $\mathit{X}$ and $\mathit{Y}$, respectively\footnote{For simplicity, we suppose the input and output domain of $p$ remain to be $X$ and $Y$, respectively.}.
MT checks whether multiple test inputs and their execution outputs satisfy an MR ($(t, t', p(t), p(t'))\in R$), whereas a conventional testing technique checks whether an individual test input's execution output is as expected ($p(t)=f(t)$).
What exactly distinguishes MT from conventional testing techniques is the concept of MR. MRs describe necessary properties that are expected to hold by a software system. Since these necessary properties are commonly derived based on a comprehensive understanding of the specification from which a software system is implemented, MRs give insight into the inner relationship among multiple software executions, and their violations indicate the inconsistency between software implementation and specification.
In this regard, MRs allow MT to examine the quality of a software system from the perspective of software's necessary properties, which is a distinctive feature of MT compared to conventional testing techniques.
Given the important role played by MRs in MT, a very natural idea is to consider the software's necessary properties as a new perspective of assessing test adequacy in addition to the aspects of program, specification, fault, and operational profile focused by conventional test adequacy criteria.

Following this viewpoint, a fundamental problem is how to reasonably take software necessary properties as a new perspective of test adequacy assessment.
Our solution is to design test adequacy criteria that specify test requirements for both MRs and source inputs based on the \emph{association relationship} established between MRs and source inputs in testing.
Specifically, MT requires the collaboration of MRs and source inputs to conduct testing, during which each MR is paired with some source inputs to construct MGs and test the SUT.
We say that an MR is \emph{associated with} a source input if they are paired together to construct an MG and test the SUT.
Considering that different MRs describe various necessary properties and different source inputs can exercise different components of SUT, which source input is associated with which MR determines which component of SUT is tested by which necessary property.
Accordingly, a very natural idea for achieving adequate testing is to \emph{impose test requirements on both MRs and source inputs according to the association relationship among them, so that MT can comprehensively exercise the SUT with the help of test inputs, and at the same time different components of SUT can be tested by different software necessary properties as much as possible}.
Intuitively speaking, for each of the MRs, the more source inputs it associates with, the higher chance it has to examine more components of the SUT, and accordingly the more adequate the testing is;
for each of the source inputs,
the more MRs it associates with, the more necessary properties are involved in testing the components of SUT, and accordingly the more adequate the testing is.

The problem lies in how we impose test requirements on MRs and source inputs.
Since existing test case adequacy criteria are well developed, we propose to employ an existing test case adequacy criterion to impose test requirements on source inputs according to the actual purpose of testing, that is, source inputs employed in MT are required to satisfy some existing test case adequacy criterion.
With regard to test requirements imposed on MRs, a natural idea is to construct a coverage domain~\cite{mathur2013foun} of SUT's necessary properties, and then to require MRs to completely cover it.
However, since it is impossible to effectively acquire all necessary properties of the SUT, and accordingly it is impossible to construct a property set for MRs to cover.
To overcome this issue, we propose to impose test requirements on MRs that help achieve ``relatively'' adequate testing.
Specifically, the MRs should be as different as possible such that each source input is associated with at least $k$ MRs that are different from each other (i.e., the expected relationship among outputs are of different form for these $k$ MRs).
The underlying idea is that:
(1) By setting the lower bound of the number of association relationship, the designed criteria satisfy finite applicability~\cite{zhu1993test} property, and accordingly, there is always a finite set of MRs and source inputs that satisfies the designed criteria;
(2) Since MRs play a critical role in test result verification,
involving multiple MRs allows the constraints used in test result verification be as different as possible, which intuitively contributes to the effectiveness of testing. It should be noted that the effectiveness of the designed test adequacy criteria directly depends on the value of $k$. \emph{An appropriate value of $k$ should be one such that increasing its value does not lead to a significant improvement in the fault detection effectiveness of MT (which will be studied in our empirical studies).}

We next present our criterion in detail, namely the $k$-MR coverage criterion.

\subsection{$k$-MR coverage criterion}

Before presenting our criterion, we first formalize some necessary notions.
Without loss of generality, we assume that an MR involves only one source input and one follow-up input. Generalizing our criterion to MRs involving multiple source and follow-up inputs is straightforward.

Given a program $p$ and its specification $s$, let $T_{s}$ denote a set of source inputs of $p$ that are generated using some existing test generation technique, $S_{\rm{MR}}$ denote a set of MRs of $p$ that are identified using some MR identification technique, and $S_{\rm{MG}}$ denote the set of MGs constructed based on the source inputs in $T_{s}$ and the MRs in $S_{\rm{MR}}$.

Our discussion starts with the following basic assumptions:
\begin{itemize}
  \item Each source input in $T_{s}$ is an eligible source test case for at least one MR in $S_{\rm{MR}}$, and each source input in $T_{s}$ is used in conjunction with at least one MR in $S_{\rm{MR}}$ to construct at least one MG.
  \item Each MR in $S_{\rm{MR}}$ is used in conjunction with at least one source input in $T_{s}$ to construct at least one MG.
  \item Each MG in $S_{\rm{MG}}$ is used to test $p$ and its corresponding execution output is verified by the MR from which it is constructed.
\end{itemize}

In order to characterize which source input is used together with which MR for testing the SUT, we present the notion of association relationship between a source input and an MR.

\begin{defi}[Association relationship between a source input and an MR]
Given a set of source inputs $T_{s}$, a set of MRs $S_{\rm{MR}}$, and a set of MGs $S_{\rm{MG}}$ constructed based on $T_{s}$ and $S_{\rm{MR}}$, association relationship is defined as a binary relation over $T_{s}$ and $S_{\rm{MR}}$, i.e., $\mathit{Coop} \subseteq T_{s} \times S_{\rm{MR}}$. Given a source input $t_{i}\in T_{s}$ and an MR $\mathit{MR}_{j}\in S_{\rm{MR}}$, $t_{i}$ is said to associate with $MR_{j}$ if (1) $\exists \mathit{MG}_{k}=\langle x_{s},x_{f}\rangle\in S_{\rm{MG}} (x_{s}=t_{i} \wedge R_{\rm{in}}(x_{s}) = x_{f})$ (where $R_{\rm{in}}$ denotes the input subrelation of $\mathit{MR}_{j}$), and (2) the execution outputs of $\mathit{MG}_{k}$ is verified by the output subrelation of $\mathit{MR}_{j}$.
\end{defi}

For simplicity, we use $\langle t_{i},\mathit{MR}_{j}\rangle \in \mathit{Coop}$ to denote $t_{i}$ is associated with $\mathit{MR}_{j}$.

Next, we formalize the test adequacy criterion for MT. Let
$\mathcal{P}$ be a set of programs,
$\mathcal{S}$ be a set of specifications,
$\mathcal{T}$ be the set of all possible inputs of the programs in $\mathcal{P}$,
$\mathcal{R}$ be the set of all possible MRs derived according to the specifications in $\mathcal{S}$,
and $2^{\mathcal{T}\times\mathcal{R}}$
be the set of all possible association relationships established over $\mathcal{T}$ and $\mathcal{R}$.

\begin{defi}[Test Adequacy Criterion for MT]
\label{def:crit}
A test adequacy criterion $\mathcal{C}_{\rm{MT}}$ for MT is defined as a function:
\begin{equation}
\mathcal{C}_{\rm{MT}}:\mathcal{P}\times\mathcal{S}\times 2^{\mathcal{T}\times\mathcal{R}}\rightarrow [0,1].
\end{equation}
\end{defi}

\noindent Suppose association relationship $\mathit{Coop}$ is established over source input set $T_{s}$ and MR set $S_{\rm{MR}}$. $\mathcal{C}_{\rm{MT}}(p,s,\mathit{Coop}) = e$ ($e\in[0,1]$) means the adequacy of testing program $p$ against specification $s$ by $\mathit{Coop}$ over $T_{s}$ and $S_{\rm{MR}}$ under criterion $\mathcal{C}_{\rm{MT}}$. In particular, when the range of $\mathcal{C}_{\rm{MT}}$ is restricted to set $\{0,1\}$ where ``0'' denotes false and ``1'' denotes true, $\mathcal{C}_{\rm{MT}}$ decides whether the association relationship $\mathit{Coop}$ over $T_{s}$ and $S_{\rm{MR}}$ is adequate for testing $p$ against $s$.

\begin{figure*}[t]
    \centering
    \includegraphics[width=1.6\columnwidth]{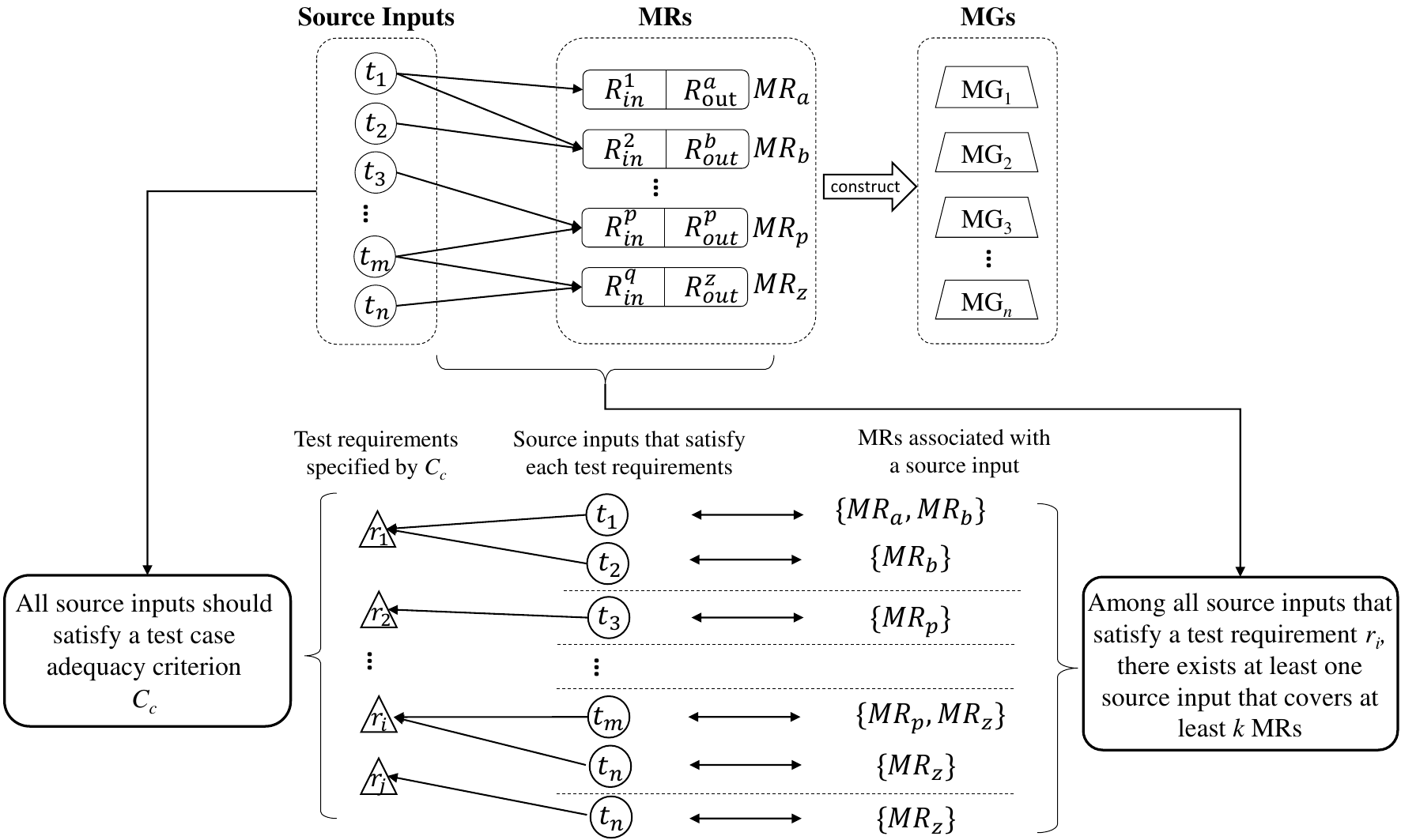}
    \caption{Illustration of $k$-MR coverage criterion. $R^{i}_{in}$ denotes the input subrelation of an MR and $R^{i}_{out}$ denotes the output subrelation of an MR.}
    \label{fig:k-cri}
\end{figure*}

Definition \ref{def:crit} involves program under test, the specification, source inputs, MRs, and the association relationship between source inputs and MRs. Accordingly, a concrete test adequacy criterion is expected to specify test requirements regarding these essential elements. Based on the above definition, we next present the $k$-MR coverage criterion, as shown in Fig.~\ref{fig:k-cri}. On the one hand, $k$-MR coverage criterion specifies test requirements for source inputs, that is, source inputs should satisfy an existing test case adequacy criterion; on the other hand, the criterion specifies test requirements for MRs and the association relationship between source inputs and MRs, which will be discussed next.

As mentioned in our motivation, we intended to encourage source inputs to associate with multiple MRs. Accordingly, we next present the notion of MRs covered by a source input.


\begin{defi}[MRs covered by a source input]
For a given source input $t$, let $S_{\rm{RO}}(t, \mathit{Coop})=\{\mathit{MR} | \langle t,\mathit{MR}\rangle\in\mathit{Coop}\}$ be the set of MRs covered by a source input.
\end{defi}

Intuitively, $S_{\rm{RO}}(t, \mathit{Coop})$ consists of all subrelations of MRs that associate with $t$.

Based on the aforementioned notions, we next give the $k$-MR coverage criterion. Given a program $p$ and its specification $s$; let $T_{s}$ and $S_{\mathit{MR}}$ be the set of source inputs and the set of MRs for conducting MT on $p$ according to $s$; let $S_{\mathit{MG}}$ be the set of MGs constructed based on $T_{s}$ and $S_{\mathit{MR}}$; let $\mathit{Coop}$ be the association relationship established over $T_{s}$ and $S_{\mathit{MR}}$ when conducting MT on $p$ using $T_{s}$, $S_{\mathit{MR}}$ and $S_{\mathit{MG}}$. Given a test case adequacy criterion $\mathcal{C}_{c}$, let $E(p,s,\mathcal{C}_{c})$ be the set of test requirements that should be satisfied to constitute adequate testing under criterion $\mathcal{C}_{c}$.

\begin{defi}[$k$-MR coverage criterion, $\mathcal{C}_{\rm{MT}}(\mathcal{C}_{c})$]
  $k$-MR coverage criterion requires that for each test requirement $r\in E(p,s,\mathcal{C}_{c})$, there exists at least one source input that satisfies it and associates with at least $k$ MRs, that is,
\begin{equation}
\begin{split}
\forall r \in & E(p,s,\mathcal{C}_{c}),\exists t \in T_{s} \\
&(sat(t,r) = true \wedge |S_{\rm{RO}}(t,\mathit{Coop})|\geq k),
\end{split}
\end{equation}
\end{defi}
in which $sat(t,r)$=\emph{true} represents that source input $t$ is able to satisfy test requirement $r$.

We next present the test adequacy measurement designed based on the $k$-MR coverage criterion.

\subsection{Test adequacy measurement}

Suppose program $p$ is tested against specification $s$ by source input set $T_{s}$ and MR set $S_{\rm{MR}}$ with association relationship $\mathit{Coop}$. Let $\mathcal{C}_{c}$ be a test case adequacy criterion that is expected to be satisfied by source inputs in $k$-MR coverage criterion, and $E(p, s, \mathcal{C}_{c})$ be the set of test requirements specified by $\mathcal{C}_{c}$ with regard to program $p$ and its specification $s$. For a given requirement $r$ in $E(p, s, \mathcal{C}_{c})$, let $sat(r, T_{s}, \mathcal{C}_{c})$ denote the subset of $T_{s}$ that can satisfy $r$ according to $\mathcal{C}_{c}$.

\begin{defi}[Test adequacy measurement based on $k$-MR coverage criterion]
On the basis of $k$-MR coverage criterion, the degree of test adequacy is calculated as follows:
\begin{equation}\label{equ:adeq}
\mathcal{C}_{\rm{MT}}^{k}(\mathcal{C}_{c}) = \frac{\sum\limits_{r\in E(p, s, \mathcal{C}_{c})} \mathcal{K}(sat(r, T_{s}, \mathcal{C}_{c}), \mathit{Coop})}{|E(p, s, \mathcal{C}_{c})|},
\end{equation}
where
\begin{equation}
\begin{split}
\mathcal{K}(T',Coop)=
\begin{cases}
  \max \limits_{t\in T'} \varepsilon(\frac{|S_{\rm{RO}}(t,Coop)|}{k}) & \mbox{if $|T'| > 0$}\\
  0 & \mbox{otherwise}
\end{cases}
,
\end{split}
\end{equation}
and
\begin{equation}
\varepsilon(n)=
\begin{cases}
  n & \mbox{if $n <1$}  \\
  1 & \mbox{otherwise}
\end{cases}
.
\end{equation}
\end{defi}

Intuitively speaking, compared with the conventional test adequacy measurement mechanism which calculates the extent to which test cases satisfy a given criterion, the proposed measurement quantifies test adequacy of MT on the basis of the key components of MT including MRs, source inputs, and the association relationship among them, which focuses on the factors that affect the test quality of MT and measures the degree of test adequacy in a more reasonable way.

It should be noted that achieving 100\% adequacy is not always possible in practice since there may be situations where identifying an MR set with at least $k$ members is difficult due to the nature of SUT.
Accordingly, under the circumstance where source test inputs have fully covered the coverage domain but a tester is unable to identify any additional MR that contributes to the adequacy of testing, MT is considered to be sufficient and can be stopped. Even so, our criterion encourages a tester to identify a set of MRs that are different from existing ones when the adequacy of MT can be improved by introducing additional MRs.

Next, we use an example to illustrate the proposed test adequacy criterion and measurement.
Consider a trigonometric function that accepts an angle value $x$ and an operation flag $f$ (whose value is either ``sine'' or ``cosine''), and computes either the sine value or cosine value of $x$ according to the value of $f$.
Let $p_{t}$ denote a program that implements the trigonometric function, $p_{t}(x,f)$ denote the output of $p_{t}$ for test input $(x,f)$. Given the following MRs identified from the trigonometric function:
\begin{itemize}
  \item[$\mathit{MR}_{1}$] For any two test inputs $(x,f)$ and $(x',f')$ such that $x'=x+360^{\circ}$ and $f'=f$, the output subrelation $R_{\rm{out}}^{1}$: $p_{t}(x,f)=p_{t}(x',f')$ must hold.
  \item[$\mathit{MR}_{2}$] For any two test inputs $(x,f)$ and $(x',f')$ such that $x'=-x$ and $f=f'=$``sine'', the output subrelation $R_{\rm{out}}^{2}$: $p_{t}(x,f)=-p_{t}(x',f')$ must hold.
  \item[$\mathit{MR}_{3}$] For any two test inputs $(x,f)$ and $(x',f')$ such that $x\in[\alpha\cdot360^{\circ}+90^{\circ}, \alpha\cdot360^{\circ} + 270^{\circ}]$ and $x'\in [\alpha\cdot360^{\circ}, \alpha\cdot360^{\circ}+180^{\circ}]$ ($\alpha\in\mathbb{Z}$, where $\mathbb{Z}$ denotes the set of integers) and $f=$``cosine'' and $f'=$``sine'', the output subrelation $R_{\rm{out}}^{3}$: $p_{t}(x,f) \leq p_{t}(x',f')$ must hold.
  \item[$\mathit{MR}_{4}$] For any two test inputs $(x,f)$ and $(x',f')$ such that $\alpha\cdot360^{\circ}\leq x \leq x' \leq \alpha\cdot360^{\circ} + 180^{\circ}$ ($\alpha\in\mathbb{Z}$) and $f=f'=$``cosine'', the output subrelation $R_{\rm{out}}^{4}$: $1 \geq p_{t}(x,f) \geq p_{t}(x',f') \geq -1$ must hold.
  \item[$\mathit{MR}_{5}$] For any two test inputs $(x,f)$ and $(x',f')$ such that $x'=x$ and $f=$``cosine'' and $f'=$``sine'', the output subrelation $R_{\rm{out}}^{5}$: $p_{t}(x,f)^{2}+p_{t}(x',f')^{2}=1$ must hold.
\end{itemize}

For simplicity, suppose $p_{t}$ has eight statements and each of them is expected to be covered in testing. That is, statement coverage (denoted by $\mathcal{C}_{stmt}$) should be satisfied; accordingly, we have eight test requirements (denoted by $r_{1}$, $r_{2}$, \dots, $r_{8}$), each of which stands for covering one of the eight statements ($s_{1}$, $s_{2}$, \dots, $s_{8}$) in $p_{t}$. Given the following source inputs $t_{1}$=($36^{\circ}$, sine), $t_{2}$=($74^{\circ}$, sine), $t_{3}$=($100^{\circ}$, cosine), $t_{4}$=($24^{\circ}$, cosine). The coverage information of these source inputs are shown in Table~\ref{tab:exmp}. For example, $t_{1}$ is able to cover $s_{1}$, $s_{2}$, and $s_{5}$; accordingly, $sat(t_{1},r_{1})$=\emph{true}, $sat(t_{1},r_{2})$=\emph{true}, $sat(t_{1},r_{5})$=\emph{true}. In addition, suppose the value of $k$ is 3.

\begin{table}[t]
    \centering
    \caption{Coverage information of $t_{1}$ to $t_{4}$ in Example 1}
    \label{tab:exmp}
    \begin{tabular}{ccccccccc}
        \toprule
        \multirow{2}*{Test input} & \multicolumn{8}{c}{Statements to cover}        \\
        \cline{2-9}
                &$s_{1}$ &$s_{2}$ &$s_{3}$ &$s_{4}$ &$s_{5}$ &$s_{6}$ &$s_{7}$ &$s_{8}$ \\
        \midrule
        $t_{1}$=($36^{\circ}$, sine) &$\surd$ &$\surd$ &        &        &$\surd$ &        &        &        \\
        $t_{2}$=($74^{\circ}$, sine) &$\surd$ &        &        &$\surd$ &        &        &$\surd$ &        \\
        $t_{3}$=($100^{\circ}$, cosine) &        &$\surd$ &$\surd$ &        &$\surd$ &        &        &        \\
        $t_{4}$=($24^{\circ}$, cosine) &        &        &$\surd$ &        &$\surd$ &$\surd$ &$\surd$ &        \\
        \bottomrule
    \end{tabular}
\end{table}

\begin{table}[t]
    \centering
    \caption{MGs constructed based on $T_{s}$ and $S_{\rm{MR}}$}
    \label{tab:exmp-mg}
    \begin{tabular}{ccc}
    \toprule
    \multirow{2}*{ID} & \multirow{2}*{MG} & Source input and MR from \\
                      &                   & which MG is constructed  \\
    \midrule
    1 & $\langle$($36^{\circ}$, sine), ($396^{\circ}$, sine)$\rangle$  & $t_{s}$: ($36^{\circ}$, sine), MR:    $\mathit{MR}_{1}$  \\
    2 & $\langle$($74^{\circ}$, sine), ($-74^{\circ}$, sine)$\rangle$  & $t_{s}$: ($74^{\circ}$, sine), MR:    $\mathit{MR}_{2}$  \\
    3 & $\langle$($100^{\circ}$, cosine), ($74^{\circ}$, sine)$\rangle$  & $t_{s}$: ($100^{\circ}$, cosine), MR: $\mathit{MR}_{3}$  \\
    4 & $\langle$($100^{\circ}$, cosine), ($124^{\circ}$,cosine)$\rangle$ & $t_{s}$: ($100^{\circ}$, cosine), MR: $\mathit{MR}_{4}$ \\
    5 & $\langle$($24^{\circ}$, cosine), ($100^{\circ}$,cosine)$\rangle$  & $t_{s}$: ($24^{\circ}$, cosine), MR: $\mathit{MR}_{4}$  \\
    6 & $\langle$($24^{\circ}$, cosine), ($24^{\circ}$,sine)$\rangle$  & $t_{s}$: ($24^{\circ}$, cosine), MR: $\mathit{MR}_{5}$  \\
    \bottomrule
    \end{tabular}
\end{table}

Consider a scenario where source input set $T_{s} = \{t_{1}, t_{2}, t_{3}, t_{4}\}$ and MR set $S_{\rm{MR}}=\{\mathit{MR}_{1}, \mathit{MR}_{2}, \mathit{MR}_{3}, \mathit{MR}_{4},$ $\mathit{MR}_{5}, \mathit{MR}_{6}\}$ are used to test $p_{t}$. A set of MGs are constructed based on $T_{s}$ and $S_{\rm{MR}}$, as shown in Table~\ref{tab:exmp-mg}. Accordingly, the association relationship established over $T_{s}$ and $S_{\rm{MR}}$ are as follows.

\begin{equation}
  \begin{split}
  \mathit{Coop}= \{&\langle t_{1},\mathit{MR}_{1}\rangle, \langle t_{2},\mathit{MR}_{2}\rangle, \langle t_{3},\mathit{MR}_{3}\rangle, \\
  &  \langle t_{3},\mathit{MR}_{4}\rangle, \langle t_{4},\mathit{MR}_{4}\rangle, \langle t_{4},\mathit{MR}_{5}\rangle \}
  \end{split}
  \end{equation}

  \begin{table}[htbp]
    \centering
    \caption{Set of MRs covered by each source input in $T_{s}$}
    \label{tab:exmp-rop}
    \begin{tabular}{cccc}
    \toprule
    Source Input & Set of MRs covered \\
    \midrule
    $t_{1}$  & \{$\mathit{MR}_{\rm{1}}$\} \\
    $t_{2}$  & \{$\mathit{MR}_{\rm{2}}$\} \\
    $t_{3}$  & \{$\mathit{MR}_{\rm{3}}$, $MR_{\rm{4}}$\} \\
    $t_{4}$  & \{$\mathit{MR}_{\rm{4}}$, $\mathit{MR}_{\rm{5}}$\} \\
    \bottomrule
    \end{tabular}
  \end{table}

To calculate the degree of test adequacy, we first summarize the set of MRs covered by each source input, as shown in Table~\ref{tab:exmp-rop}.
Next, for each statement of $p$, we figure out the set of source test inputs that can cover the statement, as shown in Table~\ref{tab:exmp-st-cov}. Finally, the degree of test adequacy is calculated as follows:

\begin{equation}\label{equ:exmp-cal}
\begin{split}
  \mathcal{C}_{\rm{MT}}^{k}(\mathcal{C}_{stmt}) &= \frac{\sum\limits_{i=1,\dots,8} \mathcal{K}(sat(s_{i}, T_{s}, \mathcal{C}_{stmt}), \mathit{Coop})}{|E(p,s,C_{stmt})|} \\
  &= \frac{\mathcal{K}_{1}+\mathcal{K}_{2}+ \dots +\mathcal{K}_{8}}{8} \\
  &= \frac{\frac{1}{3}+\frac{2}{3} + \frac{2}{3} + \frac{1}{3} + \frac{2}{3} + \frac{2}{3}+\frac{1}{3} +0}{8} \\
  &= \frac{11}{24}.
\end{split}
\end{equation}

\begin{table}[t]
    \centering
    \caption{Summary of source inputs that cover each of the statements of $p$}
    \label{tab:exmp-st-cov}
    \begin{tabular}{cccc}
    \toprule
    \multirow{2}*{Statement} & Source inputs that cover & The calculation of \\
                             & the statement ($T_{s}$)  & $\mathcal{K}(sat(r, T_{s}, \mathcal{C}_{c}), \mathit{Coop})$ \\
    \midrule
    $s_{1}$   & \{$t_1$, $t_2$\} & $\mathcal{K}_{1}=\frac{max(\{1,1\})}{3}=\frac{1}{3}$ \\
    \specialrule{0em}{1pt}{1pt}
    $s_{2}$   & \{$t_1$, $t_3$\} & $\mathcal{K}_{2}=\frac{max(\{1,2\})}{3}=\frac{2}{3}$ \\
    \specialrule{0em}{1pt}{1pt}
    $s_{3}$   & \{$t_3$, $t_4$\} & $\mathcal{K}_{3}=\frac{max(\{2,2\})}{3}=\frac{2}{3}$ \\
    \specialrule{0em}{1pt}{1pt}
    $s_{4}$   & \{$t_2$\}        & $\mathcal{K}_{4}=\frac{max(\{1\})}{3}=\frac{1}{3}$ \\
    \specialrule{0em}{1pt}{1pt}
    $s_{5}$   & \{$t_1$,$t_3$,$t_4$\} & $\mathcal{K}_{5}=\frac{max(\{1,2,2\})}{3}=\frac{2}{3}$ \\
    \specialrule{0em}{1pt}{1pt}
    $s_{6}$   & \{$t_4$\}        & $\mathcal{K}_{6}=\frac{max(\{2\})}{3}=\frac{2}{3}$ \\
    \specialrule{0em}{1pt}{1pt}
    $s_{7}$   & \{$t_2,t_4$\}    & $\mathcal{K}_{7}=\frac{max(\{1,2\})}{3}=\frac{1}{3}$ \\
    \specialrule{0em}{1pt}{1pt}
    $s_{8}$  & $\varnothing$     & $\mathcal{K}_{8}=0$ \\
    \bottomrule
    \end{tabular}
\end{table}

\begin{table*}[t]
    \centering
    \caption{Summary of Subject Programs and Generated Mutants}
    \label{tab:Summary}
   \begin{tabular}{ccccccc}
        \toprule
        \multirow{2}*{Subject system}  & \multirow{2}*{Description} & \multirow{2}*{LOC}  & \multirow{2}*{Language} & \multicolumn{3}{c}{\#Mutants}  \\
        \cline{5-7}
          &   &   &   &  \#Gen & \#Equ & \#App \\
        \midrule
        PHONE & Mobile-phone charge calculation & 107 &Java &210 &38 &172  \\
        BAGGAGE & Baggage fee charging & 101 &Java &187 &67 &120 \\
        EXPENSE & Expense reimbursement system & 117 &Java &180 &29 &151 \\
        MEAL & Meal scheduling for an airline catering provider & 150 &Java &224 &51 &173 \\
        print\_tokens & Lexical analyzer & 726 &C &697 &86 &611  \\
        print\_tokens2 & Lexical analyzer & 570 &C &851 &93 &758 \\
        grep & Regular expression processor & 10068 &C &1379 &135 &1244 \\
        \bottomrule
    \end{tabular}
\end{table*}

\section{Empirical Studies}
\label{sec:sett}
A series of empirical studies were conducted to evaluate the performance of the proposed adequacy criteria and measurement for MT.
This section describes the setting of our empirical studies.

\subsection{Research Questions}
Our empirical studies attempted to answer the following research questions:

\emph{RQ1: How effective is MT in fault detection when MT is considered as adequate with different values of $k$?}
To answer this question, we conducted experiments involving three well-known black box coverage criteria and two widely-adopted white-box coverage criteria.
In particular, we studies the effectiveness of MT with different $k$ values ($k=1,2,3,4$) and different coverage criteria, aiming to derive a reasonable value of $k$.

\emph{RQ2: On the basis of a reasonable value of $k$, how does the fault detection effectiveness of MT vary according to the degree of test adequacy?}
To answer this question, we evaluated the fault detection effectiveness of MT under different degrees of test adequacy, with the purpose of assessing how MT's fault detection effectiveness changes with the increase of the test adequacy degree.

\emph{RQ3: On the basis of a reasonable value of $k$, how does the number of MGs vary according to the degree of test adequacy?}
To answer this question, we recorded the numbers of MGs required to reach a given level of test adequacy.

\subsection{Subject Programs and Their Faulty Versions}

Our empirical studies involved seven subject programs, among which four implement real-life business workflows in different application domains (namely, PHONE, BAGGAGE, EXPENSE and MEAL), two are lexical analyzers (namely, \emph{print\_tokens} and \emph{print\_tokens2}),
and the last one is a pattern matching engine (namely \emph{grep}). Details of the subject programs, including the programming language, the lines of code, and a brief introduction, are shown in Table~\ref{tab:Summary}.
Interested readers can refer to these studies~\cite{sun2021metricplus, barus2016cost} for the detailed functional specification of the subject programs.


To evaluate the fault detection effectiveness of the proposed criterion, we used program mutation to generate faulty versions of the subject programs.
For the Java programs, we leveraged the muJava~\cite{ma2006mujava} tool to generate faulty versions. Specifically, we used all applicable method-level mutation operators (12 out of 19 method-level operators) to conduct program mutation on as many possible positions in the program source code as possible. Each of the faulty versions (aka the mutants) involved only one fault.
For the C program, we leveraged the Proteum~\cite{delamaro2001proteum} tool to generate faulty versions\footnote{Note that before generating the faulty versions of print\_tokens2, we repaired a real-life defect in its source code, which was detected in our empirical study (details given in Section 6), to eliminate its impact on the experiment.}. Similarly, we used all applicable mutation operators (66 out of 108 operators) provided by Proteum in program mutation.
Since mutation operators for C programs can produce far more numbers of changes on the source code than those operators for Java programs do, and thus a large number of mutants can be generated accordingly.
To control the number of mutants, we restricted the scope of mutation to the source code related to the key functions of the program;
in addition, we restricted that only one mutant was generated by each operator at each position that can perform mutation in the source code.
As a result, we selected 11 functions related to lexical analysis in \emph{print\_tokens}, 12 functions related to lexical analysis in \emph{print\_tokens2}, and 7 functions related to regular expression parsing and pattern matching in \emph{grep} were included in program mutation.
Also, each generated mutant involved only one fault.

Note that under certain circumstances mutants that are semantically equivalent to the original program (aka equivalent mutants) can be produced during program mutation, which should be removed from the mutant set.
However, identifying the equivalent mutants is an undecidable problem~\cite{baldwin1979heu, offutt1996detect} which requires human efforts.
To identify the equivalent mutants, we first executed all the generated mutants using a set of test cases to acquire those that produced the same outputs as the original program, and then manually checked whether the surviving mutants are equivalent to the original program.
This manual inspection is feasible due to the relatively small number of surviving mutants~\cite{yao2014study}.
The numbers of faulty versions of subject programs are shown in Table~\ref{tab:Summary}, in which ``\#Gen'', ``\#Equ'', and ``\#App'' represent the number of generated mutants, equivalent mutants, and applicable mutants, respectively.

\begin{table*}[t]
  \centering
  \caption{Details of MR identification and The size of Source Test Pool}
  \label{tab:sumMRMG}
  \begin{tabular}{cccccccc}
    \toprule
    Subject  Program & \#I-cat & \#I-cho &\#O-cat &\#O-cho &\#IO-CTF &\#MR & Size of Source Test Pool\\
    \midrule
    PHONE           & 4  & 12 & 2  & 8   & 32    & 142   & 452    \\
    BAGGAGE         & 5  & 12 & 1  & 2   & 40    & 735   & 645    \\
    EXPENSE         & 5  & 14 & 3  & 5   & 70    & 1130  & 931    \\
    MEAL            & 6  & 19 & 5  & 15  & 180   & 3512  & 2168   \\
    print\_tokens   & 7  & 31 & 11 & 70  & 2306  & 15794 & 23917  \\
    print\_tokens2  & 7  & 30 & 10 & 60  & 2484  & 22013 & 25615  \\
    grep            & 8  & 86 & 13 & 145 & 1518  & 25168 & 16362  \\
    \bottomrule
  \end{tabular}
\end{table*}

\subsection{Identification of MRs}
\label{sec:mriden}

For each of the subject programs, we used METRIC$^{+}$~\cite{sun2021metricplus} to identify the MRs from their corresponding program specification.
The basic principle of METRIC$^{+}$ is to identify an MR by comprising a sub-relation on inputs and a sub-relation on outputs which are deduced from a pair of input-and-output-based complete test frames (IO-CTFs). An IO-CTF is constructed based on the category-choice framework with distinguishing output scenarios (CHOC'LATE-DIP)~\cite{liu2015enhancing} according to the specification of a software system. It characterizes a class of inputs and their corresponding output scenario based on the concepts of I-category, I-choice, O-category, and O-choice. To identify MRs, METRIC$^{+}$ pairs IO-CTFs to form candidate pairs; for each candidate pair, the technique enables a tester to identify the possible sub-relation on outputs (denoted by $R_{out}$) according to the output test frames of IO-CTFs in the candidate pair, and then identify the possible sub-relation on inputs (denoted by $R_{in}$) according to the input test frames of IO-CTFs in the candidate pair; after that, an MR is formed by comprising $R_{in}$ and $R_{out}$. 

For each subject program, we analyzed their functional characteristics and summarized the parameters that affect the program behavior, defined the corresponding categories and choices, constructed IO-CTFs, and formed candidate pairs.
We exhausted all candidate pairs to identify as many MRs as possible for our empirical study.
Accordingly, all identified MRs formed an MR pool which is used to construct MR sets that associate with source inputs in testing.
Table~\ref{tab:sumMRMG} shows the details of MR identification for each of the subject programs, in which ``\#I-cat'', ``\#I-cho'', ``\#O-cat'', ``\#O-cho'', ``\#IO-CTF'', and ``\#MR'' represent the number of I-categories, I-choices, O-categories, O-choices, IO-CTFs, and MRs, respectively.

\subsection{Construction of Test Input Pools}

To study the performance of the proposed $k$-MR criterion when it is used in conjunction with different coverage criteria, our empirical study involved two white-box coverage criteria, including statement coverage and branch coverage, and three black-box coverage criteria, including I-choice coverage, I-choice-pair coverage, and IO-CTF coverage.
Note that the I-choice coverage criterion is adapted from the \emph{each-choice-used criterion}~\cite{ammann1994using, zhu1997software}, and the I-choice-pair coverage and the IO-CTF coverage are adapted from the \emph{all-combination criterion}~\cite{ammann1994using, zhu1997software}.

To enable the construction of source test suites according to different coverage criteria, we constructed a test pool for each subject program.
Specifically, we first constructed an initial test pool by generating test inputs based on the IO-CTFs defined for MR identification (Section~\ref{sec:mriden});
for each IO-CTF, we randomly generated 10 test inputs according to its input test frame (i.e., the valid combination of I-choices); accordingly, these test inputs enable the generation of source test suites that satisfy those black-box coverage criteria. We next analyzed the statement coverage and branch coverage of test inputs in this initial test pool using Jacoco~\cite{ecl2022jacoco} and gcov~\cite{gnu2022gcov} to determine those statements and branches that remain uncovered. After that, for the Java programs, we leveraged the EvoSuite~\cite{fraser2011evosuite} tool to generate additional inputs that cover the remaining statements and branches; the generated test inputs and the test inputs in the initial test pool collectively made up the final test pool for each of the Java subject programs.
For the C programs, since not the entire source code was involved in the generation of faulty versions, we were only interested in those test inputs that can cover the source code involved in mutant generation; accordingly, we manually analyzed the path condition of those uncovered statements and branches and designed test inputs that can cover them accordingly. The size of source test input pool for each subject program is shown in Table~\ref{tab:sumMRMG}.

\subsection{Generation of Metamorphic Test Suites}

Since MT requires not only source inputs but also MRs, we refine the notion of a test suite as a three tuple that consists of a set of source inputs $T_s$, a set of MRs $S_{\mathit{MR}}$, and the set of MG $S_{\mathit{MG}}$ generated according to $T_s$ and $S_{\mathit{MR}}$, denoted by $\langle T_s,S_{\mathit{MR}},S_{\mathit{MG}} \rangle$.
Depending on the theme of evaluation in our empirical study, we mainly generate test suites according to the following rules:
\begin{enumerate}
  \item For \emph{RQ1}, we aimed to generate test suites that fully satisfy the proposed criterion when $k$ takes different values ($k=1,2,3,4$).
      Accordingly, for a given coverage criterion $\mathcal{C}$, we first select a set of source inputs (denoted by $T_s$) that just satisfies $\mathcal{C}$, and then select a set of MRs (denoted by $S_{MR}^{k=i}$) and generate a set of MGs (denoted by $S_{MG}^{k=i}$) such that $T_s$ ,$S_{MR}^{k=i}$, and $S_{MG}^{k=i}$ together satisfy the proposed criterion when $k$ takes the value of $i$.
      Therefore, given a set of source inputs, three different sets of MRs are selected to form three test suites.
      For each coverage criterion, we constructed 100 independent sets of source inputs that satisfy the criterion, and 500 test suites were generated accordingly.
      It is worth noting that the reason we set the upper value of $k$ to 4 is that for PHONE, BAGGAGE, and grep, at most four different output subrelations can be discovered when identifying MRs.

  \item For \emph{RQ2} and \emph{RQ3}, we aimed to generate test suites achieving various test adequacy levels in a balanced manner.
      Accordingly, the value of $k$ was fixed to a specific value which is determined by the experiment result of \emph{RQ1}, and we attempted to generate test suites whose test adequacy values spread across the range of $(0,1]$.
      Specifically, we divided the value range of test adequacy into 10 disjoint but equal-size intervals, i.e., $(0,0.1],(0.1, 0.2],\dots,(0.9,1]$; each value interval is called a \emph{test adequacy level. For each test adequacy level}, we employed the simple greedy algorithm~\cite{andrew2006using, shin2018theo} to generate 100 independent test suites whose test adequacy value belongs to the interval's value range.
      That is, each test suite is generated by iteratively selecting source test inputs and MRs that contribute to the adequacy value until it is greater than the interval's lower bound and equal to or smaller than the interval's upper bound.
\end{enumerate}

\subsection{Variables and Measures}

\subsubsection{Independent variables}

The independent variables in our empirical study relate to the coverage criteria for selecting source inputs. We used five coverage criteria including statement coverage, branch coverage, I-choice coverage, I-choice-pair coverage, and IO-CTF coverage, with the target of studying the performance of the proposed $k$-MR criterion when it is used in conjunction with different test case adequacy criteria. Note that the last three criteria are derived from the \emph{each-choice-used} coverage and the \emph{all-combination} coverage in category-partition method, which is the basis for METRIC$^{+}$, the MR identification approach used in this study.

\subsubsection{Dependent Variables}

The dependent variables mainly concern about the metrics for evaluation.

For \emph{RQ1}, we used the \emph{fault detection effectiveness} (\emph{FDE}) to evaluate how effective the metamorphic test suites in terms of detecting the faulty versions. \emph{FDE} is defined as the ratio of the number of faulty versions detected by a metamorphic test suite to the total number of faulty versions. It is calculated as:
\begin{equation*}
\emph{FDE}(\emph{P}, \emph{MTS}) = \frac{\emph{N}_d}{\emph{N}_t},
\end{equation*}
where $P$ is the SUT, $\emph{MTS}$ is a metamorphic test suite, $N_d$ is the number of faulty versions detected by $\emph{MTS}$, $N_t$ is the total number of faulty versions of \emph{P}.
Apparently, the higher value of \emph{FDE} the more effective a metamorphic test suite in terms of fault detection.


For \emph{RQ2}, apart from using \emph{FDE}, we additionally used the \emph{fault detection rate} (\emph{FDR}) to evaluate the ratio of a faulty version to be revealed by metamorphic test suites selected according to the proposed criterion. For a given faulty version $v$, and a test adequacy level $l_a$, suppose we constructed $m$ metamorphic test suites whose degree of adequacy belongs to $l_a$, and among them, $q$ test suites can detect $v$, then \emph{FDR} is calculated as follows:
\begin{equation*}
\emph{FDR}(v, l_a) = \frac{q}{m}.
\end{equation*}
Apparently, the higher value of \emph{FDR} the higher rate of detecting a given fault.

For \emph{RQ3}, we recorded the number of MGs constructed when a given test adequacy level is reached.

\begin{figure}[!htp]
    \centering
    \includegraphics[width=8.8cm]{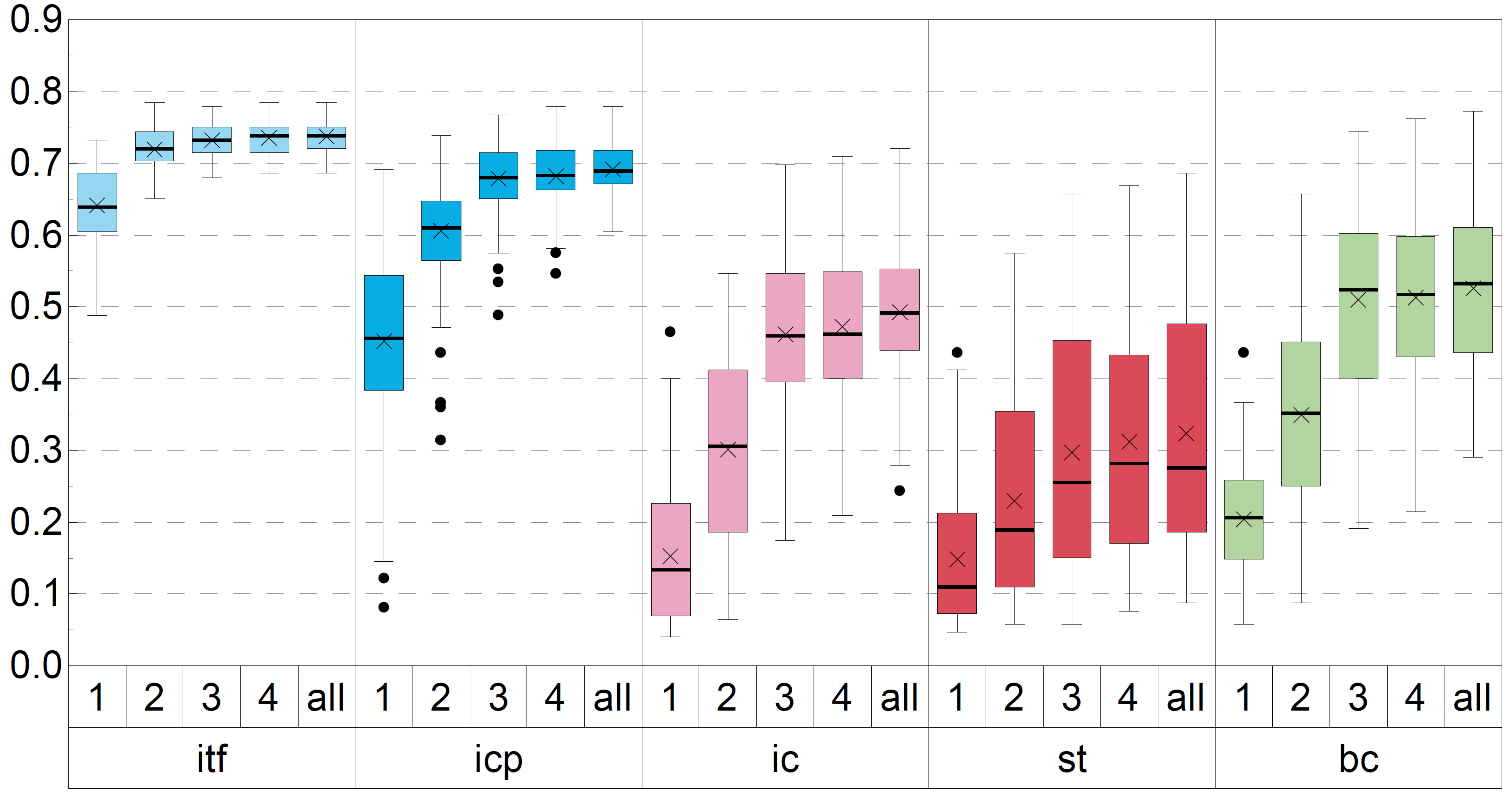}
    \caption{Distribution of the fault detection effectiveness of $k$-MR coverage criterion under different values of $k$ on program PHONE}
    \label{fig:k-ms-ph}
\end{figure}

\section{Empirical Results and Discussions}
\label{sec:empi}

\subsection{RQ1: Relationship between fault detection effectiveness and the values of $k$}
\label{sec:rq1}

One theme of our empirical study is to investigate the fault detection effectiveness of proposed test adequacy criterion under different values of $k$.
The full experimental results for this theme are given in Figure A1 in Appendix. Since all subject programs show the similar trend in their corresponding result, we only present the distribution of fault detection effectiveness of the proposed criterion for program PHONE in Figure~\ref{fig:k-ms-ph}.
In each box plot in the figure, the primary x-axis indicates the coverage criterion satisfied by source test inputs, in which ``itf'',``icp'', ``ic'', ``st'', and ``bc'' represent IO-CTF coverage, I-choice-pair coverage, I-choice coverage, statement coverage, and branch coverage, respectively; the secondary x-axis indicates the $k$ value of the proposed criterion, in which the numbers represent the values of $k$, and ``all'' represents the scenario where all applicable MRs for each source input were involved in testing. In each box, the lower and upper bounds of the box represent the first and third quartiles of distribution range, respectively. The upper and lower whiskers represent the largest and smallest data within the range of $\pm1.5\times$\emph{IQR} (i.e., the interquartile range), respectively. The middle line, cross, and solid circle represent the median value, the average value, and the outliers, respectively.

From the experimental results, we can have the following observations:
\begin{itemize}
  \item Across all subject programs and all coverage criteria, the average fault detection effectiveness increases as the value of $k$ goes up.
  \item The increment of the average fault detection effectiveness decreases as the value of $k$ increases, and the average value gradually tends to reach the upper bound, that is, the ``all'' scenario.
\end{itemize}

The first observation indicates that increasing the amount of MRs that associate with each of the source inputs can improve the fault detection effectiveness of MT. With regard to the second observation, however, it indicates that purely increasing the number of MRs can lead to diminishing marginal returns where increasing the number of MRs does not significantly yield a proportional benefit of fault detection. 
In addition, the point of diminishing marginal return differs for different coverage criteria. Specifically, for the IO-CTF coverage criterion, the distributions of fault detection effectiveness are quite similar when the value of $k$ is no less than 2; with regard to the I-choice pair coverage and the branch coverage, the distributions of fault detection effectiveness are not very much different when the value of $k$ equals to or greater than 3;
as for the I-choice coverage and statement coverage, the situation is quite similar to that of the I-choice-pair coverage and branch coverage. A plausible reason is that IO-CTF criterion is much more stronger than the other four criteria, which results in more source inputs of IO-CTF criterion than those of other criteria. Accordingly, fewer associations between MRs and source inputs are required to achieve the upper bound of fault detection effectiveness. The above observations provide a piece of intuitive evidence for the setting of $k$ in our proposed criterion. Accordingly, when the proposed criterion is used along with the IO-CTF coverage, the value of $k$ is preferred to be 2; when the criterion is used along with one of the other four coverage criteria, the value of $k$ is preferred to be 3. The follow-up experiments regarding the effectiveness of proposed criterion under different value intervals of test adequacy will be based on this set of $k$'s values.

\begin{figure*}[!htp]
    \centering
    \includegraphics[width=18cm]{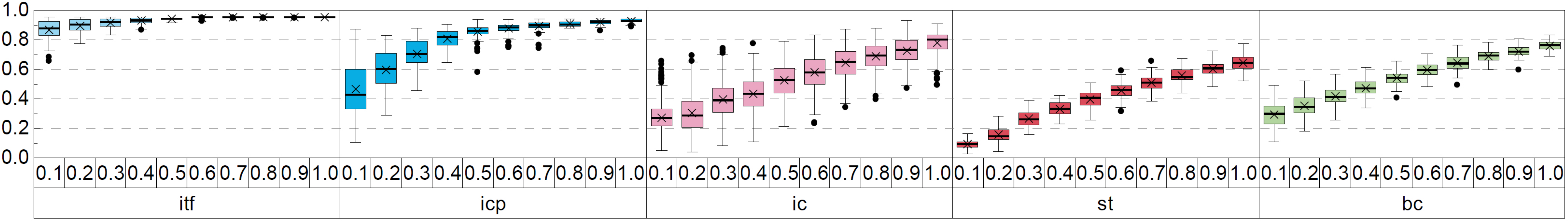}
    \caption{Distribution of the fault detection effectiveness of $k$-MR coverage criterion under different test case adequacy criteria on program \emph{print\_tokens}}
    \label{fig:cri-ms-pt}
\end{figure*}

\subsection{RQ2: Relationship between fault detection effectiveness and test adequacy levels}

This part of the experiment aims to study how MT's fault detection effectiveness changes with the variation of the test adequacy level. Accordingly, we constructed a series of test suites of different test adequacy levels. Figure~\ref{fig:cri-ms-pt} presents the distribution of fault detection effectiveness of such test suites on program \emph{print\_tokens} (full results given in Appendix). In each box plot, the primary x-axis indicates the coverage criterion satisfied by source test inputs as mentioned in Section~\ref{sec:rq1}; the secondary x-axis indicates the value intervals of test adequacy level (e.g., ``0.1'' represents the interval $(0,0.1]$).
For each interval, the box represents the fault detection effectiveness distribution of 100 independent test suites whose adequacy values belong to the interval.

Across all value intervals, when the proposed criterion was used along with the IO-CTF coverage, the resulting test suites achieve the best fault detection effectiveness among the five coverage criteria. Test suites constructed using the I-choice pair coverage generally performed the second, followed by those constructed using I-choice coverage and branch coverage, and those constructed using statement coverage performed the least. From Figure~\ref{fig:cri-ms-pt} we have the following observations:
\begin{itemize}
\item Considering all the coverage criteria, across all subject programs, the fault detection effectiveness of test suite increases with increasing the test adequacy value up to 1. Among five coverage criteria, our proposed criterion together with the IO-CTF coverage achieves the best fault detection effectiveness.
\item The distributions of fault detection effectiveness tend to change from wide regions to narrow regions for scenarios involving IO-CTF coverage and I-choice-pair coverage as the degree of test adequacy increases, while scenarios involving I-choice coverage, statement coverage, and branch coverage do not exhibit this phenomenon.
\end{itemize}

For the first observation, it indicates that achieving higher adequacy values tends to detect more faults. In addition, it is preferred to achieve a higher adequacy value when constructing a test suite based on the proposed criterion since the effectiveness of fault detection monotonically increases as the value interval goes up.
The second observation indicates that when the proposed criterion is used along with the IO-CTF coverage or the I-choice-pair coverage, the fault detection effectiveness of test suites tends to be more stable with the increment of test adequacy level.
One plausible reason for this observation is that the test suites constructed according to these coverage criteria almost revealed all faulty versions that can be revealed by the set of all identified MRs. Under such a circumstance, the number of faulty versions detected can hardly be improved furthermore; thereby the fault detection effectiveness of test suites is gradually confined to a relatively small region.

\begin{figure*}[!htp]
    \centering
    \includegraphics[height=4cm]{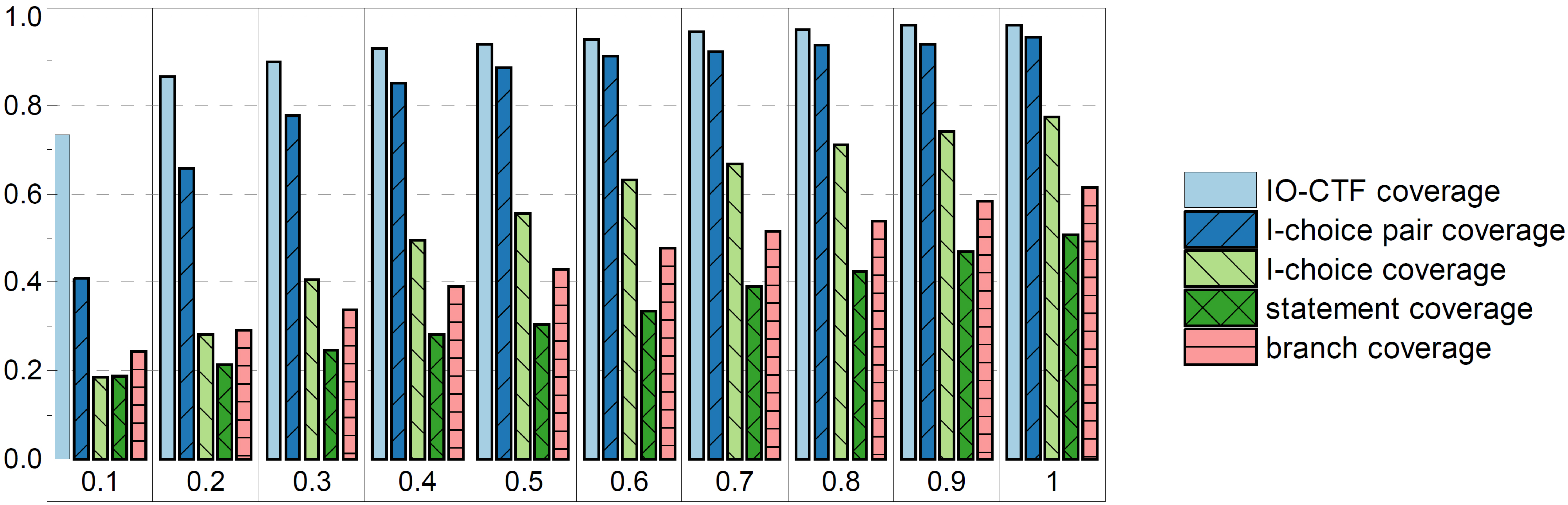}
    \caption{Average of the fault detection rate of $k$-MR coverage criterion under different test case adequacy criteria on program \emph{grep}}
    \label{fig:cri-kb-grep}
\end{figure*}

For the test suites belonging to different test adequacy levels, we also summarized their average fault detection rate with regard to all faulty versions, with an aim to study how reliable the faulty versions being detected by the test suites constructed according to our proposed criterion.
Figure~\ref{fig:cri-kb-grep} presents the average FDR values of test suites belonging to different value intervals for program \emph{grep}  (full results given in Appendix). In each bar chart, the primary x-axis indicates the value intervals of test adequacy, and the secondary x-axis indicates the coverage criterion satisfied by source test inputs. From Figure~\ref{fig:cri-kb-grep} we have the following observations:
\begin{itemize}
\item Considering all the coverage criteria, across all subject programs, test suites' average fault detection rate increases with increasing the test adequacy interval.
\item The proposed criterion together with the IO-CTF coverage achieves the best average fault detection rate. For four of the subject programs, the average fault detection rate of test suites constructed by the proposed criterion together with the IO-CTF coverage is very close to 1.
\end{itemize}

These observations are encouraging, indicating that the proposed criterion together with the IO-CTF coverage is highly reliable in terms of constructing effective test suites. With regard to the rest of the coverage criteria, I-choice-pair coverage performs the second best, followed by the I-choice coverage and branch coverage, and finally the statement coverage. In traditional testing coverage theory, branch coverage is stronger than statement coverage. Similarly, IO-CTF stronger than I-choice-pair, and I-choice-pair stronger than I-choice. Such ranking of effectiveness is consistent with the traditional test adequacy theory.

\subsection{RQ3: Number of MGs constructed under different test adequacy levels}

This part of the experiment aims to investigate how many MGs are required to achieve a given test adequacy level.
To answer this question, we summarized the number of MGs generated from test suites that satisfy different coverage criteria and test adequacy level. Figure~\ref{fig:cri-sz-exp} presents the distribution of the number of MGs for program EXPENSE (full results given in Appendix). Similarly, in each box plot, the primary x-axis indicates the coverage criterion and the secondary x-axis indicates the value intervals of test adequacy.

\begin{figure*}[!htp]
    \centering
    \includegraphics[width=15cm]{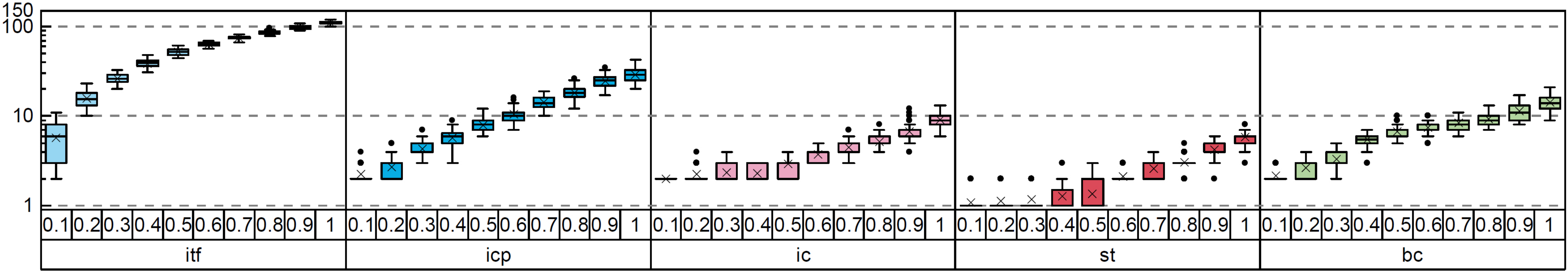}
    \caption{Distribution of the number of MGs to reach a given test adequacy level under different test case adequacy criteria.}
    \label{fig:cri-sz-exp}
\end{figure*}

Across all subjects and all value intervals, test suites constructed using the IO-CTF coverage generally produced the largest number of MGs. Test suites constructed using the I-choice-pair coverage produced the second largest number of MGs, of which the size of MG set is about one-fifth to one-tenth of that of test suites constructed using IO-CTF coverage. Moreover, test suites constructed using statement coverage generally produced the smallest number of MGs among all five criteria. This observation indicates that the number of required tests to achieve an identical adequacy value interval can vary largely for different coverage criteria. Since IO-CTF coverage subsumes~\cite{zhu1997software} the I-choice-pair coverage and the I-choice coverage, source inputs selected using IO-CTF coverage are definitely more than those selected using the latter two criteria, which accordingly leads to a larger set of MGs for the IO-CTF coverage; this situation is analogous for the branch coverage vs. the statement coverage. Traditionally, stronger coverage commonly requires more test cases while weaker coverage commonly requires fewer test cases. Our observations are consistent with this tradition.

\subsection{Threats to Validity}

In this subsection, we discuss some threats to the validity of our research outcomes.

\emph{The correctness of identified MRs.} An important issue that affects the validity of our study was the correctness of MRs identified in our experiments. The fault detection effectiveness would have been affected if incorrect MRs were used in the experiments. With regard to this issue, we thoroughly checked the correctness of each identified MRs before executing tests on subject programs.

\emph{The amount of subject program's faulty versions and their representativeness.} Obviously, the faulty versions used in our experiments directly affect the experiment results. On the one hand, the faulty versions were expected to distinguish the performance of our proposed criterion when it was used with different coverage criteria; on the other hand, the number of faulty versions was expected to be small, with an aim to control the scale of our experiments. With these considerations in mind, we employed mutation testing to generate the faulty versions. For each subject program, we employed all applicable mutation operators to generate the faulty versions of subject programs; in addition, we endeavored to perform mutation on as much source code as possible, with an aim to generate a ``diverse'' set of faulty versions. We believe that the amount of faulty versions used in the experiment is suitable for our evaluations and they are capable of representing the common faults hidden in the subject programs.

\emph{The representativeness of subject programs.} The validity of our research outcomes would be further improved if more complex subject programs were employed. On the one hand, we would like to point out that identifying MRs from subject programs is labor-intensive and time-consuming, especially for programs whose function is complex. On the other hand, we would like to point out that a vast majority of existing studies only reported a small number of MRs for various kinds of software systems, which hardly provides substantial data set that is helpful to give an insight into the test adequacy assessment issue of MT. In view of these problems, we decided to reuse the subject systems reported in~\cite{sun2021metricplus} which offered a relatively large number of MRs, and then we included three additional subject programs which have been widely adopted in existing studies~\cite{xie2013metamorphic, barus2016impact, sun2022feedback}. Previous studies has demonstrated the appropriateness of these subject programs. In addition, the selected subject programs manifest diversity in terms of both scale and application domains. Thus, we are confident that our subject programs are suitable as the benchmark programs for our empirical study. 

\emph{The correctness of subject programs.} One important issue that affects the validity of our experimental results was the correctness of subject programs. Note that our experiment revealed a real-life defect in print\_tokens2, and we corrected the source code accordingly to eliminate this defect  (details given in Section~\ref{sec:realFault}). For the remaining subject programs, we have carefully checked their source code and tested them. We ensure that the implementations of these subject programs were in line with our expectations.

\emph{Assumption of the type of MR.} Our experiment employed the METRIC$^{+}$ technique to identify MRs from the specifications of subject programs. It is worth noting that the MRs identified using METRIC$^{+}$ are of a certain type, that is, an MR can be split into an input-only relation and an output-only relation. However, not all MRs are of such type~\cite{chen2018metamorphic}. Still, it was reported that over 90\% of MRs reported in existing studies are of the aforementioned type~\cite{sun2021metricplus}, which indicates that our research outcomes can be applied to most of the testing scenarios.

\section{A Real-life Fault Detected by Our Study}
\label{sec:realFault}

\begin{figure}[t]
    \centering
    \includegraphics[width=0.99\columnwidth]{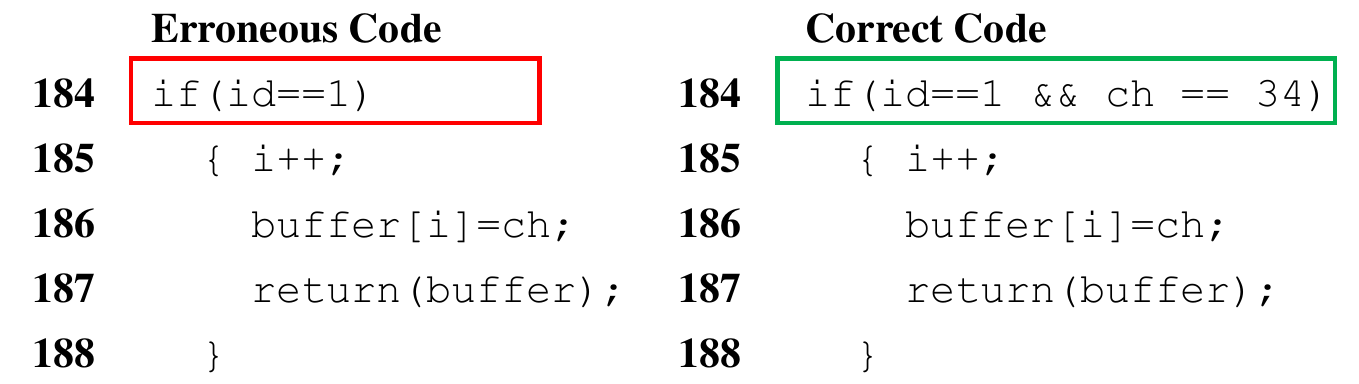}
    \caption{A real-life fault in print\_tokens2.}
    \label{fig:real-fault}
\end{figure}

In our empirical study, an unexpected defect was revealed in the program print\_tokens2, as shown in Fig.~\ref{fig:real-fault}. Specifically speaking, this defect locates at line 184 of function \texttt{get\_token(token\_stream tp)} in program print\_tokens2. This function deals with a sequence of characters from a token stream and outputs a token according to the predefined rules of syntax.
In this function, several conditional statements are used to handle different kinds of tokens, one of which is used to judge whether the terminal character of a string-type character sequence is a double quotation mark (i.e., `\textquotedbl'). However, the code of this \texttt{if} statement is ``\texttt{if (id == 1)}'' which fails to judge whether the terminal character \texttt{ch} equals to the double quotation mark. Accordingly, the program fails to correctly handle character sequences that only have a double quotation mark at the beginning of the sequence. The correct code should be ``\texttt{if (id == 1 \&\& ch == 34)}'' (in which ``34'' is the ASCII code of a double quotation mark).

This fault violates the following MR: ``Given a source input $t_s$ consisting of a string, a comma, and a number, a follow-up input $t_f$ is derived from $t_s$ by removing the second quotation mark and subsequent characters. The outputs of $t_s$ and $t_f$ are denoted as $o_s$ and $o_f$, the output tokens of $o_s$ and $o_f$ are denoted as $ot_s$ and $ot_f$, respectively. It is obvious that $ot_f$ should be the substring of $ot_s$.'' For instance, there is a source input $t_s$=<{\textquotedbl}abcd{\textquotedbl}, {\textquotesingle},{\textquotesingle}, 123>, its corresponding output is $o_s$=``string,{\textquotedbl}abcd{\textquotedbl}.{\textbackslash}ncomma.{\textbackslash}nnumeric,123.{\textbackslash}n''. Accordingly, output tokens of $o_s$ is $ot_s$=``{\textquotedbl}abcd{\textquotedbl}123''. Based on this MR, the follow-up input $t_f$=<{\textquotedbl}abcd, , > is obtained, and its output is $o_f$=``error,{\textquotedbl}{\textquotedbl}abcd{\textbackslash}n{\textquotedbl}.{\textbackslash}n''. Accordingly, output tokens of $o_f$ is $ot_f$=``{\textquotedbl}abcd{\textbackslash}n''. Clearly, $ot_f$ is not a substring of $ot_s$. Therefore, this pair of test cases makes the program violate the MR.

Although this program has been widely used as the subject program in a large number of studies, this defect has remained undetected for more than a decade regardless of the large number of testing techniques employed in these studies. This unexpected by-product of our empirical study demonstrates MT's unique capability of revealing longstanding real-life defects.

\section{Related Work}
\label{sec:rela}

In this section, we first discuss studies related to the test adequacy assessment of MT, and then discuss some recent advances in the foundational theory of MT.

\subsection{Test adequacy assessment of MT}

Assessing the test adequacy of MT has been an important research topic that deserves lots of effort to study.
So far, there only exist a few studies that follow the principle of test adequacy assessment for conventional testing techniques (e.g., to measure the coverage of test cases based on a code coverage criterion) to assess the test adequacy of MT. For example,
Ding and Hu~\cite{ding2017application} proposed a framework that evaluates the test adequacy of MT and helps guide the iterative identification of MRs and the generation of source and follow-up inputs.
They leveraged some given coverage criteria to evaluate the extent to which SUT was exercised by test inputs during testing, and new source and follow-up inputs were generated if the coverage criterion was not fully satisfied; in addition, once it was not able to improve the degree of coverage by purely generating test inputs, new MRs were identified for generating additional test inputs. In their approach, they proposed to use function coverage, statement coverage, and def-use pair coverage to evaluate the extent to which test inputs exercised the SUT. Necessary code was instrumented into the source code of SUT to acquire the coverage achieved by test inputs.
In addition, they also proposed to use mutation testing along with existing coverage criteria to further enhance the effectiveness of test inputs and MRs.
An empirical study on a Monte Carlo simulation program was conducted to evaluate the proposed framework. The empirical results showed that the coverage results of test inputs was able to indicate the quality of MT and used as clues for selecting MRs and test inputs for improving the adequacy of testing. The main difference between their framework and our criterion is that they mainly consider requirements on test inputs (that is, the coverage of test inputs) while does not pay much attention to the role of MRs; whereas our criterion specifically raises requirements on the MRs used along with the source test inputs, and accordingly the test adequacy is assessed from the test inputs and the MRs. Still, both their framework and our criterion can guide the selection of new MRs and test inputs.

Chen et al.~\cite{chen2012optimized} proposed the equivalence-class coverage criterion for every metamorphic relation (ECCEM), which combined the equivalent class testing with MT. Specifically, the input domain of SUT was first partitioned into a number of disjoint equivalent classes, and the criterion required that each MR should cover equivalent classes as many as possible.
An MR $R$ is considered to cover a given equivalent class $E$ once there is at least one MG of $R$ whose source or follow-up input belongs to $E$.
Obviously, the underlying idea of this criterion is to encourage each MR to be used with test inputs that are of different equivalent classes, which thereby enables an MR to test SUT under divergent program behaviors triggered by these test inputs.
An experiment involving a small program with seven MRs was conducted to study the proposed criterion. The experimental results showed that the MRs and test input selected based on the proposed criterion was able to achieve a high fault detection effectiveness.
In essence, this criterion shares the same purpose as ours, which is to encourage the association between MRs and test inputs. Nevertheless, this criterion achieves this purpose by carefully selecting source inputs for each MR, whereas our criterion achieves this purpose by carefully selecting MRs for each source input; in addition, this criterion assumes that each MR identified is supposed to be used in testing, which is not the case for our criterion.

Dong et al.~\cite{dong2009effective} proposed three adequacy criteria for MT based on program path coverage criterion, namely all-path coverage (APC), all-path coverage for every MR (APCEM), and all-path-pair coverage for every MR (APPCEM).
An MR is considered to be capable of covering a program path if the MR has a feasible source or follow-up input that can cover the program path.
Accordingly, APC requires that for each elementary path of SUT there exists at least one MR that covers it; APCEM requires that for each MR, all elementary paths it can cover are required to be covered in testing; APPCEM further considers the pairs of elementary paths that an MR can cover, and it requires that for each MR, all pairs of elementary paths it can cover is required to be covered in testing.
Empirical studies were conducted to evaluate the fault detection effectiveness of these criteria, and the results showed that among three criteria APCEM can achieve a relatively high fault detection effectiveness using few test inputs.
Recalling the ECCEM criterion~\cite{chen2012optimized}, APCEM and ECCEM share the same principle which is to measure the coverage of an MR by means of measuring the coverage of MR's source and follow-up inputs, whereas the coverage criterion leveraged is different. In our study, we focus on designing a criterion that specifically imposes requirements on MRs, and we have also studied our criterion when it is used along with different coverage criteria.

\subsection{Foundational Theory of MT}

MT has been receiving increasing attention in recent years due to its capability of alleviating the oracle problem~\cite{chen2018metamorphic, segura2016survey, niu2021enhance}.
Given the importance of MRs in MT, the identification of MRs is a hot research topic that has drawn much attention from the research community~\cite{altamimi2022meta}.

At present, a number of MR identification techniques have been proposed based on different intuitions: (1) pattern-based techniques~\cite{zhou2018TSE, segura2022automated} construct MRs based on some predefined template-like patterns; (2) machine learning-based techniques~\cite{kanewala2016pred, rahman2020mrpredt} identify MRs from the source code or document of SUT based on trained machine learning models; (3) category-choice-based techniques~\cite{chen2016metric, sun2021metricplus} identify MRs from the specification of SUT by analyzing pairs of complete test frames; (4) search based techniques~\cite{zhang2019auto} which identifies polynomial MRs by searching for suitable optimal parameters of some predefined parameterized relations using evolutionary algorithms; (5) the data mutation (datamorphic testing) based techniques~\cite{sun2016mumt, zhu2019datamorphic} leverage the data mutation operators to identify metamorphic relations of SUT; (6) natural language processing-based technique~\cite{blasi2021memo} derives MRs from code comments written in natural language by finding comments that may describe equality MRs and translating them into Java assertions that represents MRs. Apart from these techniques, researchers have proposed techniques that combine multiple MRs into MR compositions and a recent study has analyzed the effectiveness of MR compositions~\cite{qiu2022theoretical}.

Another important factor that strongly influences the effectiveness of MT is the source test inputs. Accordingly, researchers have proposed various techniques to generate effective source test cases: (1) constraint solving-based techniques~\cite{alatawi2016source, sun2022path} aim to generate test cases that cover different program paths; (2) adaptive random testing-based techniques~\cite{barus2016impact} generate source test cases with a high degree of diversity; (3) iterative metamorphic testing techniques~\cite{sun2020iterative} employ the follow-up inputs as new source inputs to iteratively expand the number of source inputs; (4) equivalent class based technique~\cite{chen2012optimized}. At the same time, researchers have also compared the influences of different source input generation strategies to the effectiveness of MT~\cite{chen2004meta}. In this paper, we study the test adequacy assessment of MT which is expected to provide guidelines for conducting highly effective MT when source test inputs and MRs are already available and also provide clues for generating additional source inputs and identifying additional MRs, which can help improve the test adequacy of MT.

In addition to the generation of source inputs and the identification of MRs, some studies attempted to control the use of source inputs and MRs in the testing procedure, with an aim to improve the cost-effectiveness of MT. For example, Sun et al.~\cite{sun2022feedback} proposed the feedback-directed metamorphic testing (FDMT) technique which leveraged the feedback information of each round of test execution and output verification to guide the selection of source test inputs and MRs in the following rounds of tests. Their empirical studies showed that FDMT can effectively improve the cost-effectiveness of MT. Spieker and Gotlieb~\cite{spieker2020adaptive} proposed the adaptive metamorphic testing (AMT) technique which employed reinforcement learning with contextual bandits to adaptively select MRs with high potential of fault detection effectiveness during the procedure of MT. The result of their case studies showed that AMT was useful for selecting effective MRs. Unlike these techniques, our study focuses on designing a reasonable criterion that helps to select a set of source inputs and MRs that can test the SUT as sufficiently as possible and reveal the hidden faults in SUT as effectively as possible.

\section{Conclusion and Further Work}
\label{sec:conclusion}

Metamorphic testing (MT) is highly effective in detecting software defects. However, the test adequacy assessment of MT remains an open issue that deserves effort to explore. In this paper, we have made an attempt toward this issue.
Our underlying point of view is that apart from imposing requirements on the coverage of test inputs, it is also necessary to impose requirements on the MRs used along with the test inputs.
Accordingly, we proposed a new criterion called the $k$-MR coverage criterion which specifically raises requirements on the MRs used along with the source test inputs. The proposed criterion is used together with a coverage criterion for test cases in order to assess MT's test adequacy from both the aspect of MRs and that of test inputs.
An empirical study was conducted to determine the performance of the proposed test adequacy measurement, involving seven programs and over 3000 faulty versions of them, over 68,000 MRs in total, and five coverage criteria.
The empirical results have shown that the proposed measurement can effectively indicate the fault detection effectiveness of test suites. In addition, our empirical study has revealed a real-life defect hidden in a subject program that has been extensively studied in many previous studies.

In the future, we plan to investigate the selection technique of MRs and source test inputs based on the proposed criteria, aiming to employ a subset of available MRs and source inputs to test the SUT without significantly jeopardizing the fault detection effectiveness of MT. Such a technique is expected to improve the cost-effectiveness of MT. In addition, we would like to study the prioritization technique for MRs and source inputs which orders the execution priority of MRs and source inputs such that the defects can be revealed as early as possible during testing.

\appendices
\section*{Acknowledgments}
The authors are very grateful to Professor Tsong Yueh Chen for his invaluable comments on the earlier versions of this manuscript.

\bibliographystyle{IEEEtran}
\bibliography{ref}

\begin{thebibliography}{10}
\providecommand{\url}[1]{#1}
\csname url@samestyle\endcsname
\providecommand{\newblock}{\relax}
\providecommand{\bibinfo}[2]{#2}
\providecommand{\BIBentrySTDinterwordspacing}{\spaceskip=0pt\relax}
\providecommand{\BIBentryALTinterwordstretchfactor}{4}
\providecommand{\BIBentryALTinterwordspacing}{\spaceskip=\fontdimen2\font plus
\BIBentryALTinterwordstretchfactor\fontdimen3\font minus
  \fontdimen4\font\relax}
\providecommand{\BIBforeignlanguage}[2]{{%
\expandafter\ifx\csname l@#1\endcsname\relax
\typeout{** WARNING: IEEEtran.bst: No hyphenation pattern has been}%
\typeout{** loaded for the language `#1'. Using the pattern for}%
\typeout{** the default language instead.}%
\else
\language=\csname l@#1\endcsname
\fi
#2}}
\providecommand{\BIBdecl}{\relax}
\BIBdecl

\bibitem{tricentis2017}
\BIBentryALTinterwordspacing
Tricents, ``Software fail watch: 5th edition,'' last accessed on December 7,
  2020. [Online]. Available:
  \url{https://www.tricentis.com/resources/software-fail-watch-5th-edition/}
\BIBentrySTDinterwordspacing

\bibitem{barr2015oracle}
E.~T. Barr, M.~Harman, P.~McMinn, M.~Shahbaz, and S.~Yoo, ``{The Oracle Problem
  in Software Testing: A Survey},'' \emph{IEEE Transactions on Software
  Engineering}, vol.~41, no.~5, pp. 507--525, 2015.

\bibitem{patel2018mapping}
K.~Patel and R.~M. Hierons, ``A mapping study on testing non-testable
  systems,'' \emph{Software Quality Journal}, vol.~26, no.~4, pp. 1373--1413,
  2018.

\bibitem{chen1998metamorphic}
T.~Y. Chen, S.~C. Cheung, and S.~M. Yiu, ``Metamorphic testing: a new approach
  for generating next test cases,'' Technical Report HKUST-CS98-01, Department
  of Computer Science, Hong Kong University of Science and Technology, Hong
  Kong, 1998.

\bibitem{tian2018deeptest}
Y.~Tian, K.~Pei, S.~Jana, and B.~Ray, ``Deep{T}est: automated testing of
  deep-neural-network-driven autonomous cars,'' in \emph{Proceedings of the
  40th International Conference on Software Engineering (ICSE'18)}.\hskip 1em
  plus 0.5em minus 0.4em\relax ACM, 2018, pp. 303--314.

\bibitem{xiao2022metamorphic}
D.~Xiao, Z.~Liu, Y.~Yuan, Q.~Peng, and S.~Wang, ``Metamorphic testing of deep
  learning compilers,'' \emph{Proceedings of the ACM on Measurement and
  Analysis of Computing Systems}, vol.~6, no.~1, pp. 1--28, 2022.

\bibitem{sun2014property}
C.-A. Sun, Z.~Wang, and G.~Wang, ``A property-based testing framework for
  encryption programs,'' \emph{Frontiers of Computer Science}, vol.~8, no.~3,
  pp. 478--489, 2014.

\bibitem{zhou2019TSE}
Z.~Q. Zhou, T.~H. Tse, and M.~Witheridge, ``Metamorphic robustness testing:
  Exposing hidden defects in citation statistics and journal impact factors,''
  \emph{IEEE Transactions on Software Engineering}, vol.~47, no.~6, pp.
  1164--1183, 2021.

\bibitem{yuan2021perception}
Y.~Yuan, S.~Wang, M.~Jiang, and T.~Y. Chen, ``{Perception Matters: Detecting
  Perception Failures of VQA Models Using Metamorphic Testing},'' in
  \emph{Proceedings of the IEEE/CVF Conference on Computer Vision and Pattern
  Recognition (CVPR'21)}.\hskip 1em plus 0.5em minus 0.4em\relax IEEE, 2021,
  pp. 16\,908--16\,917.

\bibitem{zhang2021deep}
Z.~Zhang, P.~Wang, H.~Guo, Z.~Wang, Y.~Zhou, and Z.~Huang, ``Deepbackground:
  Metamorphic testing for deep-learning-driven image recognition systems
  accompanied by background-relevance,'' \emph{Information and Software
  Technology}, vol. 140, pp. 106\,701:1--106\,701:14, 2021.

\bibitem{sun2024ConMT}
C.-A. Sun, H.~Dai, N.~Geng, H.~Liu, T.~Y. Chen, P.~Wu, Y.~Cai, and J.~Wang,
  ``An interleaving-guided metamorphic testing approach for concurrent
  programs,'' \emph{ACM Transactions on Softwware Engineering and Methodology},
  vol.~33, no.~1, pp. 8:1--8:21, 2024.

\bibitem{IEEE2021Standard}
{ISO/IEC/IEEE}, ``{ISO/IEC/IEEE International Standard -- Software and systems
  engineering -- Software testing -- Part 4: Test techniques},''
  \emph{{ISO/IEC/IEEE 29119-4:2021(E)}}, pp. 1--148, 2021.

\bibitem{rahamn2021using}
K.~Rahman, I.~Kahanda, and U.~Kanewala, ``{MRpredT: Using Text Mining for
  Metamorphic Relation Prediction},'' in \emph{Proceedings of the IEEE/ACM 42nd
  International Conference on Software Engineering Workshops (ICSEW'21)}.\hskip
  1em plus 0.5em minus 0.4em\relax ACM, 2021, pp. 420--424.

\bibitem{qiu2022theoretical}
K.~Qiu, Z.~Zheng, T.~Y. Chen, and P.-L. Poon, ``Theoretical and empirical
  analyses of the effectiveness of metamorphic relation composition,''
  \emph{IEEE Transactions on Software Engineering}, vol.~48, no.~3, pp.
  1001--1017, 2022.

\bibitem{sun2016mumt}
C.-A. Sun, Y.~Liu, Z.~Wang, and W.~Chan, ``$\mu${MT}: A data mutation directed
  metamorphic relation acquisition methodology,'' in \emph{Proceedings of the
  1st International Workshop on Metamorphic Testing, in conjunction with the
  38th International Conference on Software Engineering (ICSE'16)}.\hskip 1em
  plus 0.5em minus 0.4em\relax ACM, 2016, pp. 12--18.

\bibitem{chen2016metric}
T.~Y. Chen, P.-L. Poon, and X.~Xie, ``M{ETRIC}: {MET}amorphic {R}elation
  {I}dentification based on the {C}ategory-choice framework,'' \emph{Journal of
  Systems and Software}, vol. 116, pp. 177--190, 2016.

\bibitem{sun2021metricplus}
C.-A. Sun, A.~Fu, P.-L. Poon, X.~Xie, H.~Liu, and T.~Y. Chen,
  ``{METRIC\textsuperscript{+}: A Metamorphic Relation Identification Technique
  Based on Input plus Output Domains},'' \emph{IEEE Transactions on Software
  Engineering}, vol.~47, no.~9, pp. 1764--1785, 2021.

\bibitem{segura2022automated}
S.~Segura, J.~C. Alonso, A.~Martin-Lopez, A.~Dur{\'a}n, J.~Troya, and
  A.~Ruiz-Cort{\'e}s, ``Automated generation of metamorphic relations for
  query-based systems,'' in \emph{Proceedings of the 7th International Workshop
  on Metamorphic Testing, in conjunction with the 44th International Conference
  on Software Engineering (ICSE'22)}.\hskip 1em plus 0.5em minus 0.4em\relax
  ACM, 2022, pp. 48--55.

\bibitem{liu2014effectively}
H.~Liu, F.-C. Kuo, D.~Towey, and T.~Y. Chen, ``How effectively does metamorphic
  testing alleviate the oracle problem?'' \emph{IEEE Transactions on Software
  Engineering}, vol.~40, no.~1, pp. 4--22, 2014.

\bibitem{mayer2006empirical}
J.~Mayer and R.~Guderlei, ``An empirical study on the selection of good
  metamorphic relations,'' in \emph{Proceedings of the 30th Annual
  International Computer Software and Applications Conference (COMPSAC'06)},
  vol.~1.\hskip 1em plus 0.5em minus 0.4em\relax IEEE, 2006, pp. 475--484.

\bibitem{sun2022path}
C.-A. Sun, B.~Liu, A.~Fu, Y.~Liu, and H.~Liu, ``Path-directed source test case
  generation and prioritization in metamorphic testing,'' \emph{Journal of
  Systems and Software}, vol. 183, pp. 111\,091:1--111\,091:14, 2022.

\bibitem{alatawi2016source}
E.~Alatawi, T.~Miller, and H.~S\o{}ndergaard, ``Generating source inputs for
  metamorphic testing using dynamic symbolic execution,'' in \emph{Proceedings
  of the 1st International Workshop on Metamorphic Testing (MET'16), Co-located
  with the 38th International Conference on Software Engineering
  (ICSE'16)}.\hskip 1em plus 0.5em minus 0.4em\relax ACM, 2016, pp. 19--25.

\bibitem{barus2016impact}
A.~C. Barus, T.~Y. Chen, F.-C. Kuo, H.~Liu, and H.~W. Schmidt, ``The impact of
  source test case selection on the effectiveness of metamorphic testing,'' in
  \emph{Proceedings of the 1st International Workshop on Metamorphic Testing
  (MET'16), Co-located with the 38th International Conference on Software
  Engineering (ICSE'16)}.\hskip 1em plus 0.5em minus 0.4em\relax ACM, 2016, pp.
  5--11.

\bibitem{saha2018fault}
P.~Saha and U.~Kanewala, ``Fault detection effectiveness of source test case
  generation strategies for metamorphic testing,'' in \emph{Proceedings of the
  3rd International Workshop on Metamorphic Testing (MET'18), Co-located with
  the 40th International Conference on Software Engineering (ICSE'18)}.\hskip
  1em plus 0.5em minus 0.4em\relax ACM, 2018, pp. 2--9.

\bibitem{zhu1997software}
H.~Zhu, P.~A.~V. Hall, and J.~H.~R. May, ``Software unit test coverage and
  adequacy,'' \emph{ACM Computing Surveys}, vol.~29, no.~4, pp. 366--427, 1997.

\bibitem{ding2017application}
J.~Ding and X.-H. Hu, ``{Application of metamorphic testing monitored by test
  adequacy in a Monte Carlo simulation program},'' \emph{Software Quality
  Journal}, vol.~25, no.~3, pp. 841--869, 2017.

\bibitem{lascu2021dreaming}
A.~Lascu, M.~Windsor, A.~F. Donaldson, T.~Grosser, and J.~Wickerson, ``Dreaming
  up metamorphic relations: Experiences from three fuzzer tools,'' in
  \emph{Proceedings of the 6th International Workshop on Metamorphic Testing
  (MET'21), Co-located with the 43th International Conference on Software
  Engineering (ICSE'21)}.\hskip 1em plus 0.5em minus 0.4em\relax IEEE, 2021,
  pp. 61--68.

\bibitem{mathur2013foun}
A.~P. Mathur, ``Test adequacy basics,'' in \emph{Foundations of Software
  Testing, Second Edition}.\hskip 1em plus 0.5em minus 0.4em\relax Pearson
  Education India, 2013.

\bibitem{bernot1991software}
G.~Bernot, M.~C. Gaudel, and B.~Maree, ``Software testing based on formal
  specifications: a theory and a tool,'' \emph{Software Engineering Journal},
  vol.~6, no.~6, pp. 387--405, 1991.

\bibitem{demillo1989test}
R.~A. DeMillo, ``Test adequacy and program mutation,'' in \emph{Proceedings of
  the 11th international conference on Software engineering (ICSE'89)}.\hskip
  1em plus 0.5em minus 0.4em\relax ACM, 1989, pp. 355--356.

\bibitem{miranda2017ass}
B.~Miranda and A.~Bertolino, ``An assessment of operational coverage as both an
  adequacy and a selection criterion for operational profile based testing,''
  \emph{Software Quality Journal}, vol.~26, no.~4, pp. 1571--1594, 2017.

\bibitem{chen2018metamorphic}
T.~Y. Chen, F.-C. Kuo, H.~Liu, P.-L. Poon, D.~Towey, T.~Tse, and Z.~Q. Zhou,
  ``Metamorphic testing: A review of challenges and opportunities,'' \emph{ACM
  Computing Surveys}, vol.~51, no.~1, pp. 4:1--4:27, 2018.

\bibitem{poole1975debugging}
P.~C. Poole, ``Debugging and testing,'' in \emph{{Software Engineering: An
  Advanced Course}}.\hskip 1em plus 0.5em minus 0.4em\relax Berlin, Heidelberg:
  Springer, 1975, pp. 278--318.

\bibitem{goodenough1975toward}
J.~B. Goodenough and S.~L. Gerhart, ``Toward a theory of test data selection,''
  \emph{IEEE Transactions on Software Engineering}, vol. SE-1, no.~2, pp.
  156--173, 1975.

\bibitem{weyuker1986axio}
E.~J. Weyuker, ``Axiomatizing software test data adequacy,'' \emph{IEEE
  Transactions on Software Engineering}, vol. SE-12, no.~12, pp. 1128--1138,
  1986.

\bibitem{zhu1993test}
H.~Zhu and P.~A.~V. Hall, ``Test data adequacy measurement,'' \emph{Software
  Engineering Journal}, vol.~8, no.~1, pp. 21--30, 1993.

\bibitem{demillo1978hints}
R.~DeMillo, R.~Lipton, and F.~Sayward, ``{Hints on Test Data Selection: Help
  for the Practicing Programmer},'' \emph{Computer}, vol.~11, no.~4, pp.
  34--41, 1978.

\bibitem{frankl1988app}
P.~G. Frankl and E.~J. Weyuker, ``An applicable family of data flow testing
  criteria,'' \emph{IEEE Transactions on Software Engineering}, vol.~14,
  no.~10, pp. 1483--1498, 1988.

\bibitem{shi2019industry}
H.~Shi, R.~Wang, Y.~Fu, M.~Wang, X.~Shi, X.~Jiao, H.~Song, Y.~Jiang, and
  J.~Sun, ``Industry practice of coverage-guided enterprise linux kernel
  fuzzing,'' in \emph{Proceedings of the 2019 27th ACM Joint Meeting on
  European Software Engineering Conference and Symposium on the Foundations of
  Software Engineering (ESEC/FSE'19)}.\hskip 1em plus 0.5em minus 0.4em\relax
  ACM, 2019, pp. 986--995.

\bibitem{bogner2021industry}
J.~Bogner, J.~Fritzsch, S.~Wagner, and A.~Zimmermann, ``Industry practices and
  challenges for the evolvability assurance of microservices,'' \emph{Empirical
  Software Engineering}, vol.~26, no.~5, pp. 104:1--104:39, 2021.

\bibitem{barus2016cost}
A.~C. Barus, T.~Y. Chen, F.-C. Kuo, H.~Liu, R.~Merkel, and G.~Rothermel, ``A
  cost-effective random testing method for programs with non-numeric inputs,''
  \emph{IEEE Transactions on Computers}, vol.~65, no.~12, pp. 3509--3523, 2016.

\bibitem{ma2006mujava}
Y.-S. Ma, J.~Offutt, and Y.-R. Kwon, ``{muJava}: a mutation system for java,''
  in \emph{Proceedings of the 28th International Conference on Software
  Engineering (ICSE 2006)}.\hskip 1em plus 0.5em minus 0.4em\relax ACM, 2006,
  pp. 827--830.

\bibitem{delamaro2001proteum}
M.~E. Delamaro, J.~C. Maldonado, and A.~M.~R. Vincenzi, ``{Proteum/IM 2.0: An
  integrated mutation testing environment},'' in \emph{Mutation testing for the
  new century}.\hskip 1em plus 0.5em minus 0.4em\relax Springer, 2001, pp.
  91--101.

\bibitem{baldwin1979heu}
D.~Baldwin and F.~Sayward, ``Heuristics for determining equivalence of program
  mutations,'' Research Report \#161, Department of Computer Science, Yale
  University, New Haven, 1979.

\bibitem{offutt1996detect}
J.~Offutt and J.~Pan, ``Detecting equivalent mutants and the feasible path
  problem,'' in \emph{Proceedings of 11th Annual Conference on Computer
  Assurance (COMPASS'96)}.\hskip 1em plus 0.5em minus 0.4em\relax IEEE, 1996,
  pp. 17--21.

\bibitem{yao2014study}
X.~Yao, M.~Harman, and Y.~Jia, ``{A Study of Equivalent and Stubborn Mutation
  Operators Using Human Analysis of Equivalence},'' in \emph{Proceedings of the
  36th International Conference on Software Engineering (ICSE'14)}.\hskip 1em
  plus 0.5em minus 0.4em\relax IEEE, 2014, pp. 919--930.

\bibitem{liu2015enhancing}
H.~Liu, P.-L. Poon, and T.~Y. Chen, ``Enhancing partition testing through
  output variation,'' in \emph{Proceedings of the 37th International Conference
  on Software Engineering (ICSE'15)}.\hskip 1em plus 0.5em minus 0.4em\relax
  IEEE, 2015, pp. 805--806.

\bibitem{ammann1994using}
P.~Ammann and J.~Offutt, ``Using formal methods to derive test frames in
  category-partition testing,'' in \emph{Proceedings of the 9th Annual
  Conference on Computer Assurance (COMPASS'94)}.\hskip 1em plus 0.5em minus
  0.4em\relax IEEE, 1994, pp. 69--79.

\bibitem{ecl2022jacoco}
\BIBentryALTinterwordspacing
EclEmma, ``Jacoco java code coverage library,'' last accessed on June 20, 2022.
  [Online]. Available: \url{https://www.eclemma.org/jacoco/}
\BIBentrySTDinterwordspacing

\bibitem{gnu2022gcov}
\BIBentryALTinterwordspacing
GNU, ``gcov---a test coverage program,'' last accessed on June 20, 2022.
  [Online]. Available: \url{https://gcc.gnu.org/onlinedocs/gcc/Gcov.html}
\BIBentrySTDinterwordspacing

\bibitem{fraser2011evosuite}
G.~Fraser and A.~Arcuri, ``Evosuite: Automatic test suite generation for
  object-oriented software,'' in \emph{Proceedings of the 19th ACM SIGSOFT
  Symposium and the 13th European Conference on Foundations of Software
  Engineering (ESEC/FSE'11)}.\hskip 1em plus 0.5em minus 0.4em\relax ACM, 2011,
  pp. 416--419.

\bibitem{andrew2006using}
J.~Andrews, L.~Briand, Y.~Labiche, and A.~Namin, ``Using mutation analysis for
  assessing and comparing testing coverage criteria,'' \emph{IEEE Transactions
  on Software Engineering}, vol.~32, no.~8, pp. 608--624, 2006.

\bibitem{shin2018theo}
D.~Shin, S.~Yoo, and D.-H. Bae, ``A theoretical and empirical study of
  diversity-aware mutation adequacy criterion,'' \emph{IEEE Transactions on
  Software Engineering}, vol.~44, no.~10, pp. 914--931, 2018.

\bibitem{xie2013metamorphic}
X.~Xie, W.~E. Wong, T.~Y. Chen, and B.~Xu, ``Metamorphic slice: an application
  in spectrum-based fault localization,'' \emph{Information and Software
  Technology}, vol.~55, no.~5, pp. 866--879, 2013.

\bibitem{sun2022feedback}
C.-A. Sun, H.~Dai, H.~Liu, and T.~Y. Chen, ``Feedback-directed metamorphic
  testing,'' \emph{ACM Transactions on Softwware Engineering and Methodology},
  vol.~32, no.~1, pp. 20:1--20:34, 2023.

\bibitem{chen2012optimized}
L.~Chen, L.~Cai, J.~Liu, Z.~Liu, S.~Wei, and P.~Liu, ``An optimized method for
  generating cases of metamorphic testing,'' in \emph{Proceedings of the 6th
  International Conference on New Trends in Information Science, Service
  Science and Data Mining (ISSDM'12)}.\hskip 1em plus 0.5em minus 0.4em\relax
  IEEE, 2012, pp. 439--443.

\bibitem{dong2009effective}
G.~Dong, C.~Nie, and B.~Xu, ``Effectively metamorphic testing based on program
  path analysis,'' \emph{Chinese Journal of Computers}, vol.~32, no.~5, pp.
  1002--1013, 2009.

\bibitem{segura2016survey}
S.~Segura, G.~Fraser, A.~B. Sanchez, and A.~Ruiz-Cort{\'e}s, ``A survey on
  metamorphic testing,'' \emph{IEEE Transactions on Software Engineering},
  vol.~42, no.~9, pp. 805--824, 2016.

\bibitem{niu2021enhance}
X.~Niu, Y.~Sun, H.~Wu, G.~Li, C.~Nie, L.~Yu, and X.~Wang, ``Enhance
  combinatorial testing with metamorphic relations,'' \emph{IEEE Transactions
  on Software Engineering}, vol.~48, no.~12, pp. 5007--5029, 2022.

\bibitem{altamimi2022meta}
E.~Altamimi, A.~Elkawakjy, and C.~Catal, ``Metamorphic relation automation:
  Rationale, challenges, and solution directions,'' \emph{Journal of Software:
  Evolution and Process}, vol.~35, no.~1, p. e2509, 2023.

\bibitem{zhou2018TSE}
Z.~Q. Zhou, L.~Sun, T.~Y. Chen, and D.~Towey, ``Metamorphic relations for
  enhancing system understanding and use,'' \emph{IEEE Transactions on Software
  Engineering}, vol.~46, no.~10, pp. 1120--1154, 2020.

\bibitem{kanewala2016pred}
U.~Kanewala, J.~M. Bieman, and A.~Ben-Hur, ``Predicting metamorphic relations
  for testing scientific software: a machine learning approach using graph
  kernels,'' \emph{Software Testing, Verification and Reliability}, vol.~26,
  no.~3, pp. 245--269, 2016.

\bibitem{rahman2020mrpredt}
K.~Rahman, I.~Kahanda, and U.~Kanewala, ``Mrpredt: Using text mining for
  metamorphic relation prediction,'' in \emph{Proceedings of the IEEE/ACM 42nd
  International Conference on Software Engineering Workshops}.\hskip 1em plus
  0.5em minus 0.4em\relax ACM, 2020, pp. 420--424.

\bibitem{zhang2019auto}
B.~Zhang, H.~Zhang, J.~Chen, D.~Hao, and P.~Moscato, ``Automatic discovery and
  cleansing of numerical metamorphic relations,'' in \emph{Proceedings of the
  35th IEEE International Conference on Software Maintenance and Evolution
  (ICSME'19)}.\hskip 1em plus 0.5em minus 0.4em\relax IEEE, 2019, pp. 235--245.

\bibitem{zhu2019datamorphic}
H.~Zhu, D.~Liu, I.~Bayley, R.~Harrison, and F.~Cuzzolin, ``Datamorphic testing:
  A method for testing intelligent applications,'' in \emph{Proceedings of the
  1st IEEE International Conference On Artificial Intelligence Testing
  (AITest'19)}.\hskip 1em plus 0.5em minus 0.4em\relax IEEE, 2019, pp.
  149--156.

\bibitem{blasi2021memo}
A.~Blasi, A.~Gorla, M.~D. Ernst, M.~Pezz{\`e}, and A.~Carzaniga, ``{MeMo:
  Automatically identifying metamorphic relations in Javadoc comments for test
  automation},'' \emph{Journal of Systems and Software}, vol. 181, pp.
  111\,041:1--111\,041:13, 2021.

\bibitem{sun2020iterative}
C.-A. Sun, A.~Fu, Y.~Liu, Q.~Wen, Z.~Wang, P.~Wu, and T.~Y. Chen, ``An
  iterative metamorphic testing technique for web services and case studies,''
  \emph{International Journal of Web and Grid Services}, vol.~16, no.~4, pp.
  364--392, 2020.

\bibitem{chen2004meta}
T.~Y. Chen, F.-C. Kuo, Y.~Liu, and A.~Tang, ``Metamorphic testing and testing
  with special values,'' in \emph{Proceedings of the 5th ACIS International
  Conference on Software Engineering, Artificial Intelligence, Networking and
  Parallel/Distributed Computing (SNPD'04)}, 2004, pp. 128--134.

\bibitem{spieker2020adaptive}
H.~Spieker and A.~Gotlieb, ``Adaptive metamorphic testing with contextual
  bandits,'' \emph{Journal of Systems and Software}, vol. 165, pp.
  110\,574:1--110\,574:14, 2020.

\end{thebibliography}

\begin{IEEEbiography}[{\includegraphics[width=1in,height=1.25in,clip,keepaspectratio]{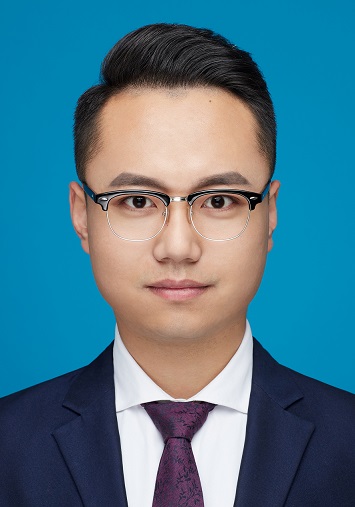}}]{An Fu}
is a PhD student in the School of Computer and Communication Engineering, University of Science and Technology Beijing, China. He received the bachelor degree in Information Security from University of Science and Technology Beijing, China. His current research interests include software testing and debugging.
\end{IEEEbiography}

\begin{IEEEbiography}[{\includegraphics[width=1in,height=1.25in,clip,keepaspectratio]{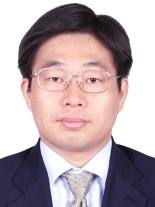}}]{Chang-ai Sun}
is a Professor in the School of Computer and Communication Engineering, University of Science and Technology Beijing. Before that, he was an Assistant Professor at Beijing Jiaotong University, China, a postdoctoral fellow at the Swinburne University of Technology, Australia, and a postdoctoral fellow at the University of Groningen, The Netherlands. He received the bachelor degree in Computer Science from the University of Science and Technology Beijing, China, and the PhD degree in Computer Science from the Beihang University, China. His research interests include software testing, program analysis, and Service-Oriented Computing.
\end{IEEEbiography}

\begin{IEEEbiography}[{\includegraphics[width=1in,height=1.25in,clip,keepaspectratio]{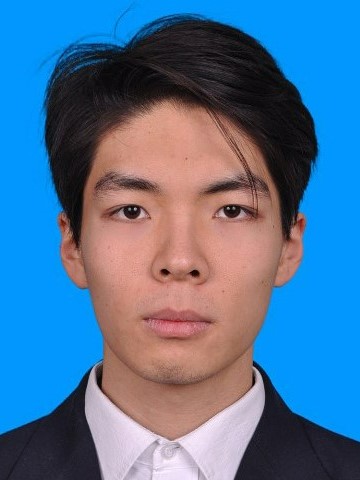}}]{Jiaming Zhang}
is a PhD student in the School of Computer and Communication Engineering, University of Science and Technology Beijing, China. He received the bachelor degree in Software Engineering and the master degree in Computer Science both from Beijing Information Science and Technology University, China. His current research interests include software testing and debugging.
\end{IEEEbiography}

\begin{IEEEbiography}[{\includegraphics[width=1in,height=1.25in,clip,keepaspectratio]{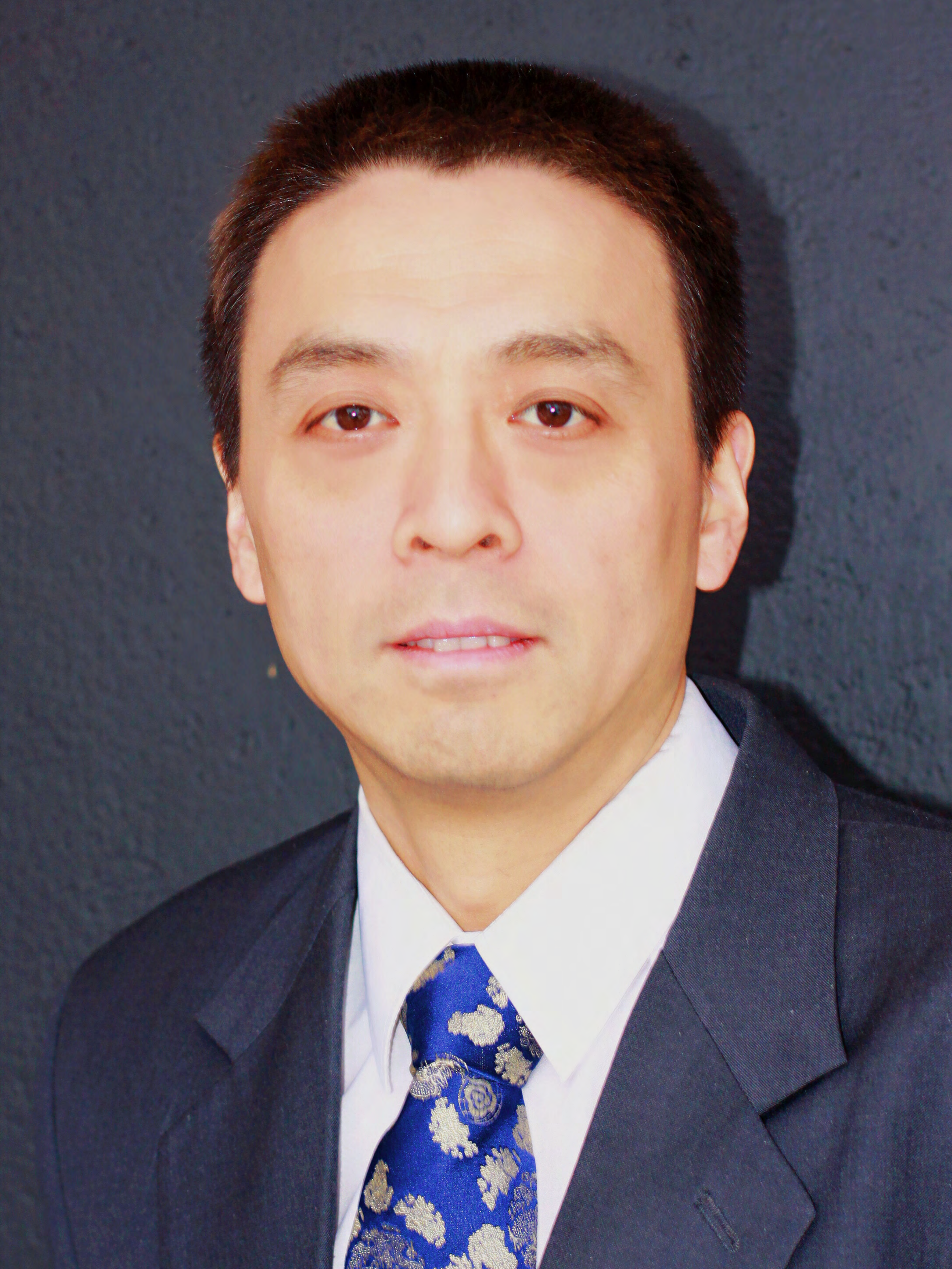}}]{Huai Liu}
is a Senior Lecturer in the Department of Computing Technologies at Swinburne University of Technology, Melbourne, Australia. He has worked as a Lecturer at Victoria University and a Research Fellow at RMIT University. He received the BEng in Physioelectronic Technology and MEng in Communications and Information Systems, both from Nankai University, China, and the PhD degree in Software Engineering from the Swinburne University of Technology, Australia. His current research interests include software testing, cloud computing, and end-user software engineering. He is a senior member of the IEEE.
\end{IEEEbiography}

\vfill

\begin{onecolumn}
\appendix[Full Experimental Results]
\label{app:fullResults}

\begin{figure*}[!htp]
    \centering
    \includegraphics[width=18cm]{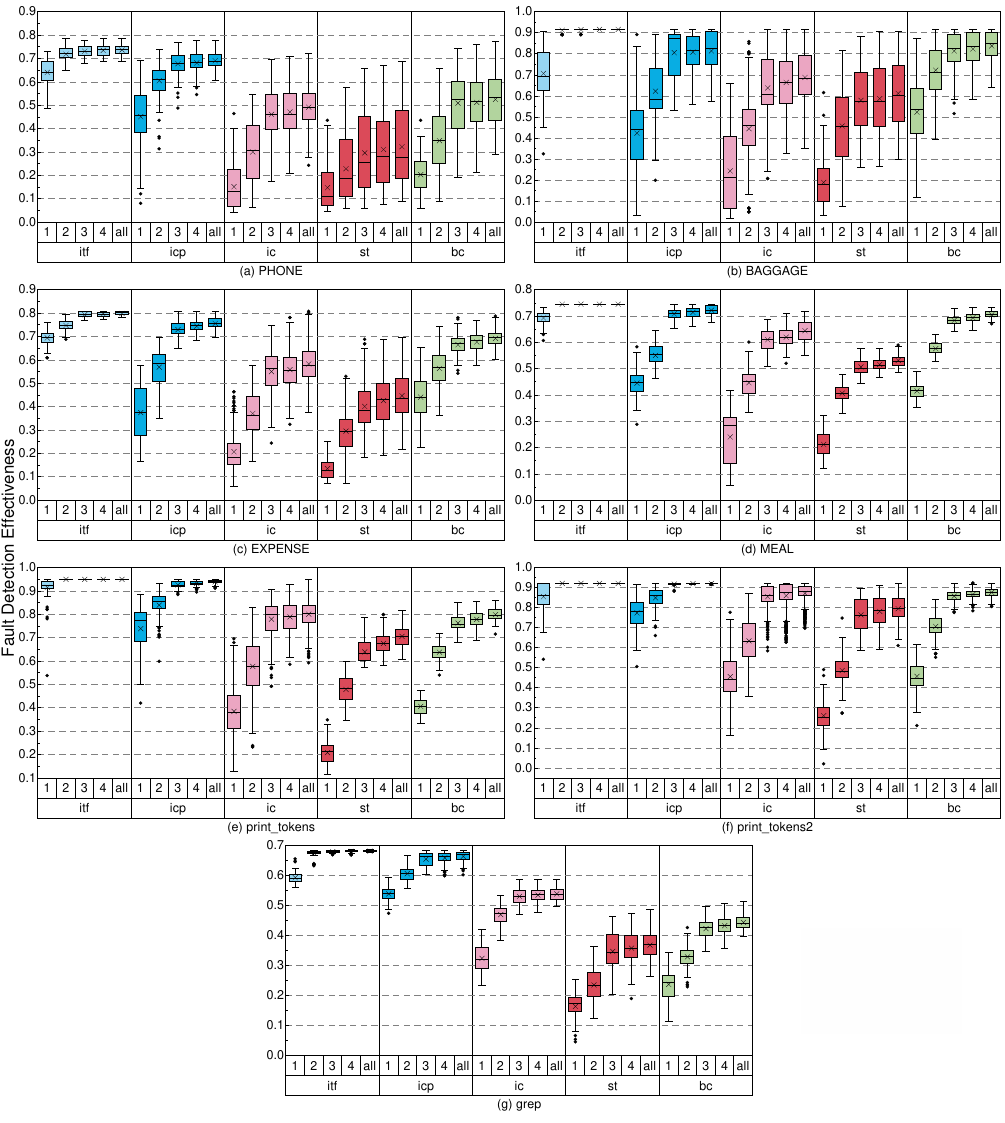}
    \caption{Distribution of the fault detection effectiveness of $k$-MR coverage criterion under different values of $k$}
    \label{fig:all-k-ms}
\end{figure*}

\begin{figure*}[!htp]
    \centering
    \includegraphics[width=18cm]{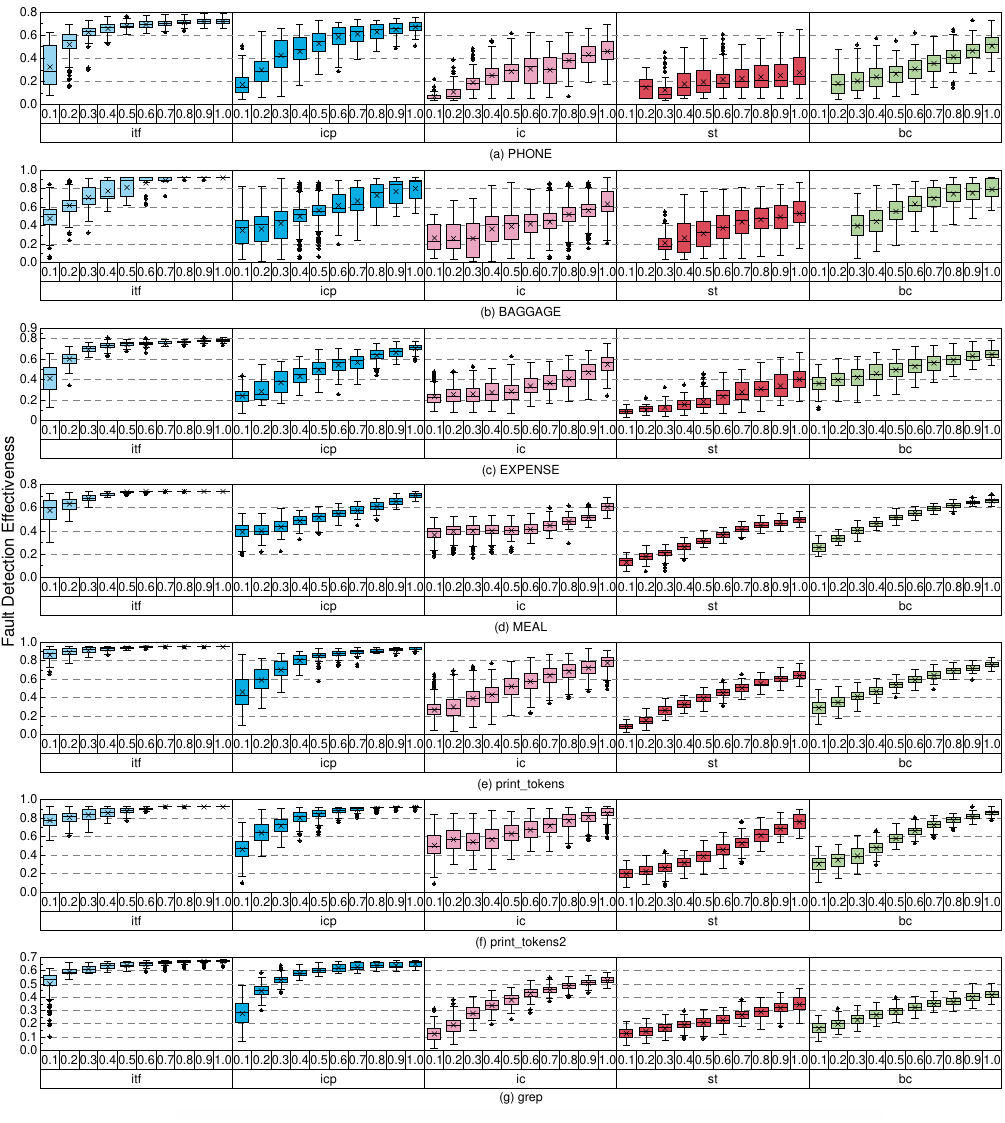}
    \caption{Distribution of the fault detection effectiveness of $k$-MR coverage criterion under different test case adequacy criteria}
    \label{fig:all-cri-ms}
\end{figure*}

\begin{figure*}[!htp]
    \centering
    \includegraphics[width=18cm]{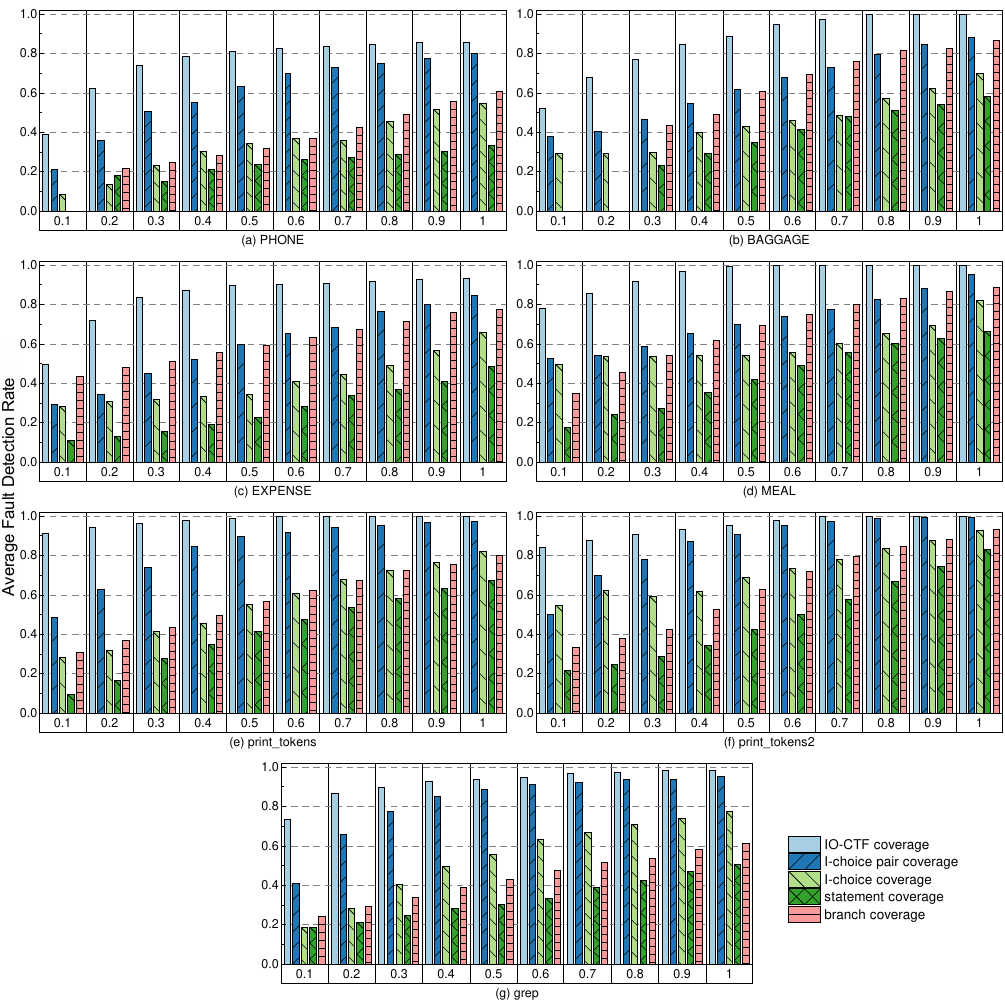}
    \caption{Average of the fault detection rate of $k$-MR coverage criterion under different test case adequacy criteria}
    \label{fig:all-cri-kb}
\end{figure*}

\begin{figure*}[!htp]
    \centering
    \includegraphics[width=17.6cm]{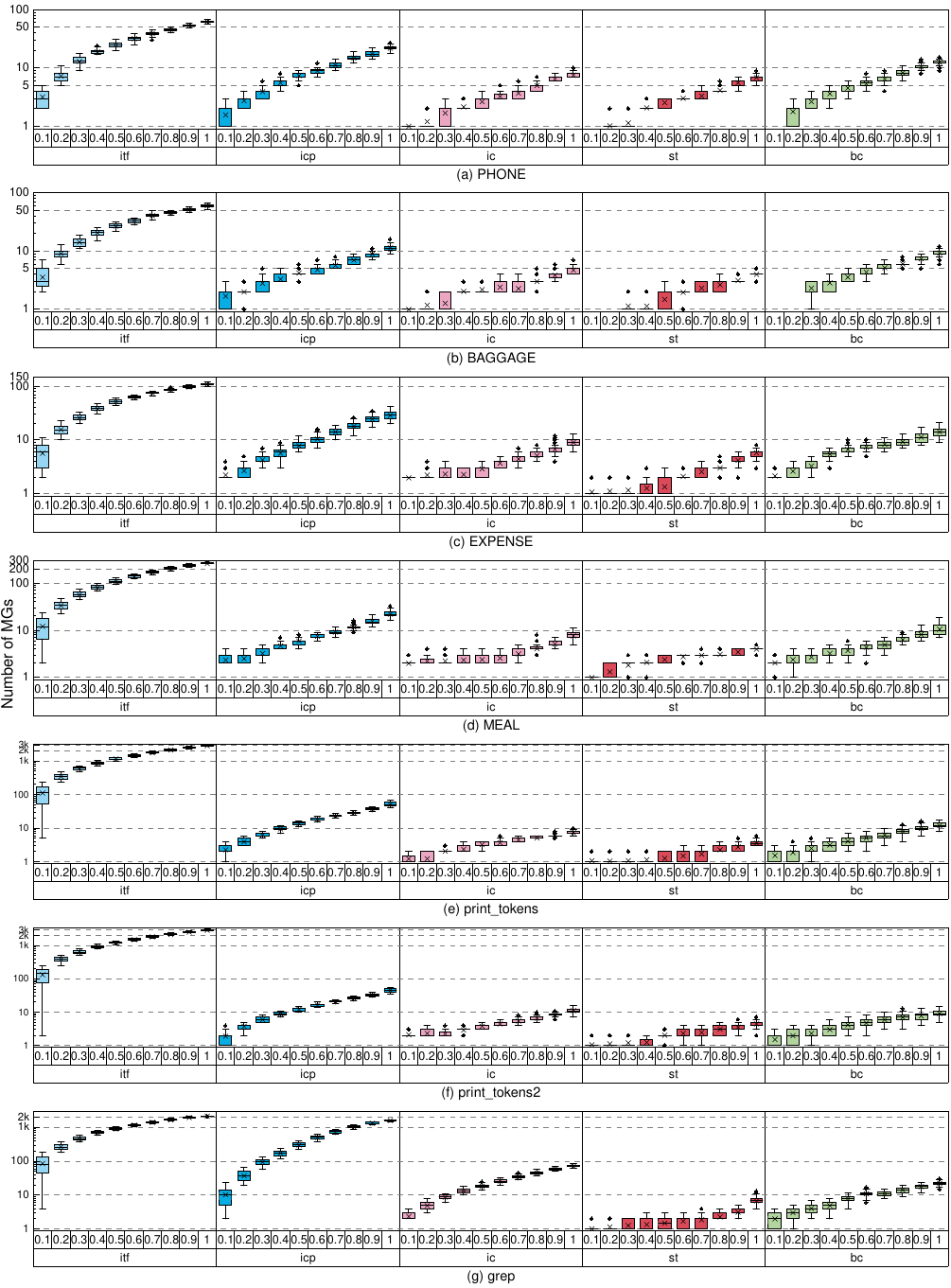}
    \caption{Distribution of the number of MGs to reach a given test adequacy level under different test case adequacy criteria}
    \label{fig:all-cri-sz}
\end{figure*}
\end{onecolumn}

\end{document}